\documentstyle[12pt,epsbox]{article}
\newcommand {\m}{\mu}
\newcommand {\n}{\nu}
\newcommand {\pl}{\partial}

\newcommand {\vp}{\varphi}

\newcommand {\al}{\alpha}
\newcommand {\be}{\beta}
\newcommand {\ga}{\gamma}

\newcommand {\la}{\lambda}

\newcommand {\si}{\sigma}

\newcommand {\del}  {\delta}
\newcommand {\Del}  {\Delta}
\newcommand {\mn}{{\mu\nu}}
\newcommand {\ls}   {{\lambda\sigma}}
\newcommand {\ab}   {{\alpha\beta}}
\newcommand {\half}{ {\frac{1}{2}} }

\newcommand {\change} {\leftrightarrow}

\newcommand {\ul}   {\underline}
\newcommand {\pr}   {{\quad .}}
\newcommand {\com}  {{\quad ,}}
\newcommand {\q}    {\quad}

\newcommand {\qqq}   {\quad\quad\quad}
\newcommand {\qqqq}   {\quad\quad\quad\quad}

\newcommand {\lb}    {\linebreak}
\newcommand {\nl}    {\newline}
\newcommand {\nn}    {\nonumber}
\newcommand {\vs}[1]  { \vspace*{#1 cm} }

\newcounter{eq}
\newcounter{sc}

\newcommand {\MPL}  {Mod.Phys.Lett.}
\newcommand {\NP}   {Nucl.Phys.}
\newcommand {\PL}   {Phys.Lett.}
\newcommand {\PR}   {Phys.Rev.}

\newcommand {\CQG}   {Class.Quantum Grav.}
\begin{document}

%%%%%%%%%%%%%%%%%%%%%%%%%%%%%%%%%%%%%%%%%%%%%%%%%%%%%%%%%%%%%%%%%%
%%%%%%%%%%%%%%%%%%%%%%%% Title %%%%%%%%%%%%%%%%%%%%%%%%%%%%%%%%%%%
%%%%%%%%%%%%%%%%%%%%%%%%%%%%%%%%%%%%%%%%%%%%%%%%%%%%%%%%%%%%%%%%%%

\begin{flushright}
US-96-05\\
July,1996\\
hep-th/9609014
\end{flushright}
\vspace{24pt}

%\magnification=\magstep1
\pagestyle{empty}
\baselineskip15pt
%\font\cmssB=cmss17
%\font\cmssS=cmss10

\begin{center}
{\large\bf 
New Algorithm for\\
Tensor Calculation in Field Theories
\vskip 1mm
}
\vspace{10mm}
          Shoichi ICHINOSE
          \footnote{ E-mail address:\ ichinose@u-shizuoka-ken.ac.jp}\\
\vspace{5mm}
          Department of Physics, Universuty of Shizuoka,\\ 
          Yada 52-1, Shizuoka 422, Japan
\end{center}

\vspace{15mm}
\begin{abstract}
Tensor calculation of suffix-contraction  is
carried out by a C-program. Tensors are represented graphically, 
and the algorithm makes use of the topology of the graphs.
Classical and quantum gravity, in the weak-field
perturbative approach, is a special interest.
Examples of the leading order calculation of 
some general invariants such as $R_{\mn\ls}R^{\mn\ls}$~
are given.
Application to  Weyl anomaly calculation is commented.
\vspace{15mm}

%{\it PACS No}:\ 02.70.+d;\ 04.20.-q;\ 04.60.+n;\ 11.10.-z\nl
{\it Keywords}:\ Computer algebra;\ General relativity;\ Tensor calculation;\ 
Suffix contraction;\ Graphical representation;\ Topology of graph

\end{abstract}

\newpage
\pagestyle{plain}
\pagenumbering{arabic}

%%%%%%%%%%%%%%%%%%%%%%%%%%%%%%%%%%%%%%%%%%%%%%%%%%%%%%%%%%%%%%%%%%
%%%%%%%%%%%%%%%%%%%%%%%% Article %%%%%%%%%%%%%%%%%%%%%%%%%%%%%%%%%
%%%%%%%%%%%%%%%%%%%%%%%%%%%%%%%%%%%%%%%%%%%%%%%%%%%%%%%%%%%%%%%%%%

%%%%%%%%%%%%%%%%%%%%%%%%%%%%%%%%%%%%%%%%%%%%%%%%%%%%%%%%%%%%%%%%%%%%%
%%%%%%%%%%%%%%%%%%%%%%%%%%%%%%   SEC  1    %%%%%%%%%%%%%%%%%%%%%%%%%%
%%%%%%%%%%%%%%%%%%%%%%%%%%%%%%%%%%%%%%%%%%%%%%%%%%%%%%%%%%%%%%%%%%%%%
\section{Introduction}
The field theory 
is described by a Lagrangian which has fields of
different spins and their different  derivatives:\ 
$\vp,\ \pl_\m\vp,\ \pl_\m\pl_\n\vp,\ A_\m,\ \pl_\n A_\m,\ h_\mn,\ 
\pl_\al h_\mn,\ \cdots$\ 
($\vp$~:~scalar,\ $A_\m$~:~vector,\ $h_\mn$~:~(2nd rank) tensor\ )
.\ 
For the n-dimensional flat Euclidean(Lorentzian) space,
a general form is a k-th rank tensor ( we call it
{\it k-tensor} in the following, k=0,1,$\cdots$) under the global SO(n)
(SO(n-1,1)) symmetry.
In the electro-magnetism,
the energy density is given by, in the Euclidean space, an SO(n)-invariant:
%****(intro.1)%%%%%%%%%%%%%%%%%%%%
\begin{eqnarray}
{\vec E}^2+{\vec B}^2 &=& \half F_{\mn}F_{\mn} \nn\\
                      &=& (\pl_\m A_\n)^2-\pl_\m A_\n\cdot\pl_\n A_\m
                                \com\label{intro.1}\\
     F_\mn &=& \pl_\m A_\n-\pl_\n A_\m\pr\nn
\end{eqnarray}
%%%%%%%%%%%%%%%%%%%%%%%%%%%%%%
Similar thing can be said for the Yang-Mills theory.
For the curved space, it is valid 
when we treat a gravitational theory
in the weak-field perturbation~:~$g_\mn=\del_\mn+h_\mn, |h_\mn|\ll 1$.
The scalar curvature $R$ is given as
%****(intro.2)%%%%%%%%%%%%%%%%%%%%
\begin{eqnarray}
R=\pl_\m\pl_\m h_{\n\n}-\pl_\m\pl_\n h_\mn+O(h^2)
                                \pr\label{intro.2}
\end{eqnarray}
%%%%%%%%%%%%%%%%%%%%%%%%%%%%%%
This is a 0-tensor.
Generally physical quantities are expressed by 
invariants (0-tensors) and all suffixes (both derivative-suffixes
and field-suffixes) are {\it contracted}
\footnote{Generally the summation over all space-suffixes which makes
an inner product, such as 
$A_\m A_\m\equiv \sum_{\m=1}^{n}A_\m A_\m$, is called Einstein's
{\it contraction}.
         }
. 
It is sometimes, however, cumbersome to do calculation involving
the suffix-contraction, especially
for a higher-spin fields such as gravitational ones. 

The importance of high-computational ability in the development of
the quantum gravity is 
historically shown as follows. In 1963 Feynman\cite{F} calculated some
scattering amplitudes involving gravitons (,which turned out
to be an important finding in  gauge theories: Feynman-DeWitt-
Faddeev-Popov ghost) by doing algebraic computer calculation.
In 1974, 'tHooft and Veltman\cite{tHV} 
fully computed the 1-loop ( both matter-field loop and graviton loop) 
counter-terms
of gravitational theories for the first time (by 'finger' calculation
using a highly-refined calculational method called 'background-field
method'). This result means the quantum gravity is (1-loop) unrenormalizable
when matter fields are coupled. It urged theoretical physicists, 
in the latter half of 70's, to look for
or examine other alternative gravitational models such as supergravity.
In 1985, Goroff and Sagnotti\cite{GS} succeeded in calculating the full
2-loop counter-terms of the pure gravity using a C-language program.
Their result was re-confirmed by van de Ven\cite{vdV} in 1992 by use of
a nice algebraic software called 'FORM'\cite{Ver}. 
The gauge dependence of their results was examined by  \cite{SI92}
using a REDUCE program.
The result that
2-loop counter-terms do not vanish on-shell let many people have an opinion
that Einstein gravity could not be regarded as the fundamental quantum theory.
The present high activity in the string theory is based on these 
past experiences.
The above history of 
these three and half decays clearly shows, for 
important development of gravity, good algebraic computer programs 
and  efficient computational techniques
play  important roles.

This paper is partly motivated by the analysis of the Weyl anomaly.
Compared with the chiral anomaly, the general structure of the Weyl anomaly
is not so simple. ( For the chiral one, there is a result
for the general n dimension \cite{BZ}.) There exist some 
analysis \cite{BPB86,DS93,AKMM95,KMM96} and the
most popular belief is that they are given by, at each (even) dimension, 
the Euler term, some Weyl invariants and some 'trivial terms'.
The explicit results are obtained only up to the 4 dimensional case
\cite{BD}. Clearly a way to improve the situation 
is to explicitly
obtain the Weyl anomaly in higher dimensions. We tried it at the 6 dim, but
we soon realize its calculation involves so much tensor contraction that is
difficult in the hand-calculation. Furthermore, the popular algebraic softwares
are  inconvenient or inefficient for the problem
\footnote{
Most softwares so far treat suffixes directly and the permutation
symmetry w.r.t. suffixes is coded by explicitly writing (or declaring)
the equality between quantities with the permuted suffixes. This
algorithm is not appropriate in the higher dimensions or in higher
orders. See Sec.6.
}
.

In this paper, 
we present a new algorithm for the tensor contraction and present
a C-program ( we call it ``WEAKGRAV'') using it. 
Important tensors often have high permutation
symmetries with respect to their suffixes. For example, the Riemann tensor
has
%****(intro.3)%%%%%%%%%%%%%%%%%%%%
\begin{eqnarray}
R_{\mn\ab} = R_{\ab\mn}=-R_{\n\m\ab}=-R_{\mn\be\al}\pr\label{intro.3}
\end{eqnarray}
%%%%%%%%%%%%%%%%%%%%%%%%%%%%%%
When we treat 'products' of these tensors, it is most efficient to
express the tensors by  graphs which have their permutation symmetries\cite{SI}.
All tensor products are completely specified by the {\it topology} of graphs. 
It makes one
free from bothering about suffixes. 
We have only to deal with {\it topology} of graphs, {\it not} their {\it suffixes}.
The present algorithm makes use of this idea. 
In \cite{SI}, the general tensors in the curved space are treated.
In the present paper, we treat global SO(n) tensors in the flat space.

Generally the local properties of the quantum gravity can be analysed by
the perturbation around the flat space. For example, the Weyl anomaly
, the effective action and the counterterms are obtained. 
( We use, in the final form, the fact that they must be invariant
under the general-coordinate transformation. This procedure allows us 
to take into account all orders of the weak-field expansion.)
In the case,
there appear many terms of ``products'' of global SO(n) tensors.
The present algorithm is useful in such cases. 

The main purpose of this paper is to explain how the above idea about 
tensor-contraction is realized in the explicit computer calculation.
We give two sample calculations. 
Emphasis is not on the calculated results themselves but on 
how the results are obtained. 
Such presentation is good for readers to apply the present algorithm to their own
problems. 

Finally we describe the main features of the program ``WEAKGRAV''.
\begin{itemize}
\item
{\it Computers}:\ all computers running C-program
\item
{\it Program language used}:\ C-language
\item
{\it No. of lines including sample data}:\ 300
\item
{\it Input files needed}:\ indata.dat
\item
{\it Purpose of program}:\ In the perturbative treatment of 
gravitational theories,
we face many terms with complicated suffix contraction. This program helps
to do such calculation very fast. Each term is graphically 
represented. Topological
indices of graphs are utilized in this algorithm. It can be applied
to general field theories.
\item
{\it Restrictions on the complexity of the problem}:\ 
The program is at a basic level. 
The contraction terms are restricted to the quadratic type ones:\ 
$\pl_\m\pl_\n h_\ls\cdot\pl_\al\pl_\be h_{\ga\del}$. This is,
however, sufficient to do Weyl anomaly calculation of 4 dim quantum gravity.
\end{itemize}
The program ``WEAKGRAV'' is available on the request to the
author's home institute. ( It is restricted only to non-commercial use.) 

The content is organized in the following way. In Sec.2 we explain
the graphical representation of global SO(n) tensors, which is
used in the explanation of the algorithm. The codes for the tensors
are explained in Sec.3. In Sec.4, new quantities called ``{\it indices}''
are introduced and explained. They express the topology of
global SO(n) invariants. We demonstrate the calculation
using two simple examples in Sec.5. We conclude in Sec.6.
In App.A, all $(\pl\pl h)^2$-invariants are graphically
shown. Finally, in App.B, we add the input and output
data for the sample calculations of Sec.5.
%%%%%%%%%%%%%%%%%%%%%%%%%%%%%%%%%%%%%%%%%%%%%%%%%%%%%%%%%%%%%%%%%%%%%
%%%%%%%%%%%%%%%%%%%%%%%%%%%%%%   SEC  2    %%%%%%%%%%%%%%%%%%%%%%%%%%
%%%%%%%%%%%%%%%%%%%%%%%%%%%%%%%%%%%%%%%%%%%%%%%%%%%%%%%%%%%%%%%%%%%%%
\section{Graphical Representation of Global SO(n) Tensors}
%In the programming, we  code  a tensor in the form of a multi-dim array.
In the programming explanation, we use a graphical representation 
for a tensor because it can clearly express the 
connectivity of suffixes in a tensor\cite{SI,II}. 
In this section we explain 
the minimum amount of the representation necessary for the present 
programming.

We can graphically express a basic element of the present program:\ 
4-tensor\ 
$\pl_\m\pl_\n h_\ab$\ as in Fig.1.
%%%%%%%%%%%%%%%%%%%%%%%% Fig.1 %%%%%%%%%%%%%%%%%%%%%%%%%%%%%%%%%
\begin{figure}
     \centerline{
%{\epsfxsize=220pt  \epsfysize=50pt \epsfbox{P1.eps}}
\psbox[width=50mm,height=30mm]{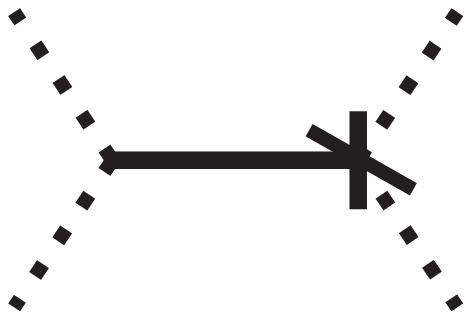}
                 }
\caption{
4-tensor $\pl_\m\pl_\n h_\ab$
        }
\label{fig1}
\end{figure}
%%%%%%%%%%%%%%%%%%%%%%%%%%%%%%%%%%%%%%%%%%%%%%%%%%%%%%%%%%%%%%%%%%%%%
We call the dotted lines {\it suffix-lines}, the rigid line {\it bond} and
the intersections of two dotted lines and one bond {\it vertices}.
There are two kinds of vertices:\ the vertex with a 
crossing symbol '$\times$' and
that one without it. 
The suffixes of derivatives ($\m,\n$\ in Fig.1) are associated with
those suffix-lines which extend from a vertex {\it without} 
the crossing symbol,
whereas the suffixes of fields ($\al,\be$\ in Fig.1) are associated
with those which extend from a vertex {\it with} it.
This graph respects all symmetries of $\pl_\m\pl_\n h_\ab$\ w.r.t. 
the permutation of suffixes:\ 
$\pl_\m\pl_\n h_\ab=\pl_\n\pl_\m h_\ab=\pl_\m\pl_\n h_{\be\al}$.
The suffix {\it contraction} is expressed by connecting the same suffix-lines. 
For example, 2-tensors\ :\ 
$\pl^2h_\ab\ ,\ \pl_\m\pl_\n h_{\al\al}\ ,\ \pl_\m\pl_\be h_\ab\ $
, which are made from Fig.1 by connecting two suffix-lines, are expressed as
in Fig.2.
%\newpage

%%%%%%%%%%%%%%%%%%%%%%%% Fig.2 %%%%%%%%%%%%%%%%%%%%%%%%%%%%%%%%%
\begin{figure}
     \centerline{
\psbox[width=150mm,height=40mm]{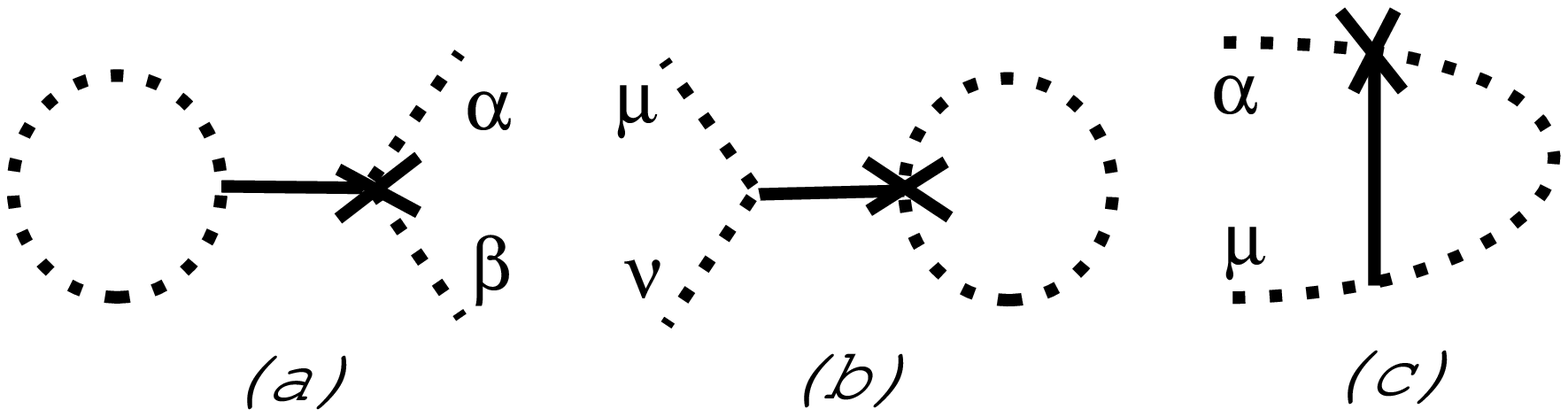}
                 }
\caption{
2-tensors of \ (a)\ 
$\pl^2h_\ab\ ,\ (b)\ \pl_\m\pl_\n h_{\al\al}\ $ and \ (c)\ 
$\pl_\m\pl_\be h_\ab\ $
        }
\label{fig2}
\end{figure}
%%%%%%%%%%%%%%%%%%%%%%%%%%%%%%%%%%%%%%%%%%%%%%%%%%%%%%%%%%%%%%%%%%%%%
%%%%%%%%%%%%%%%%%%%%%%%% Fig.3 %%%%%%%%%%%%%%%%%%%%%%%%%%%%%%%%%
\begin{figure}
     \centerline{
\psbox[width=120mm,height=50mm]{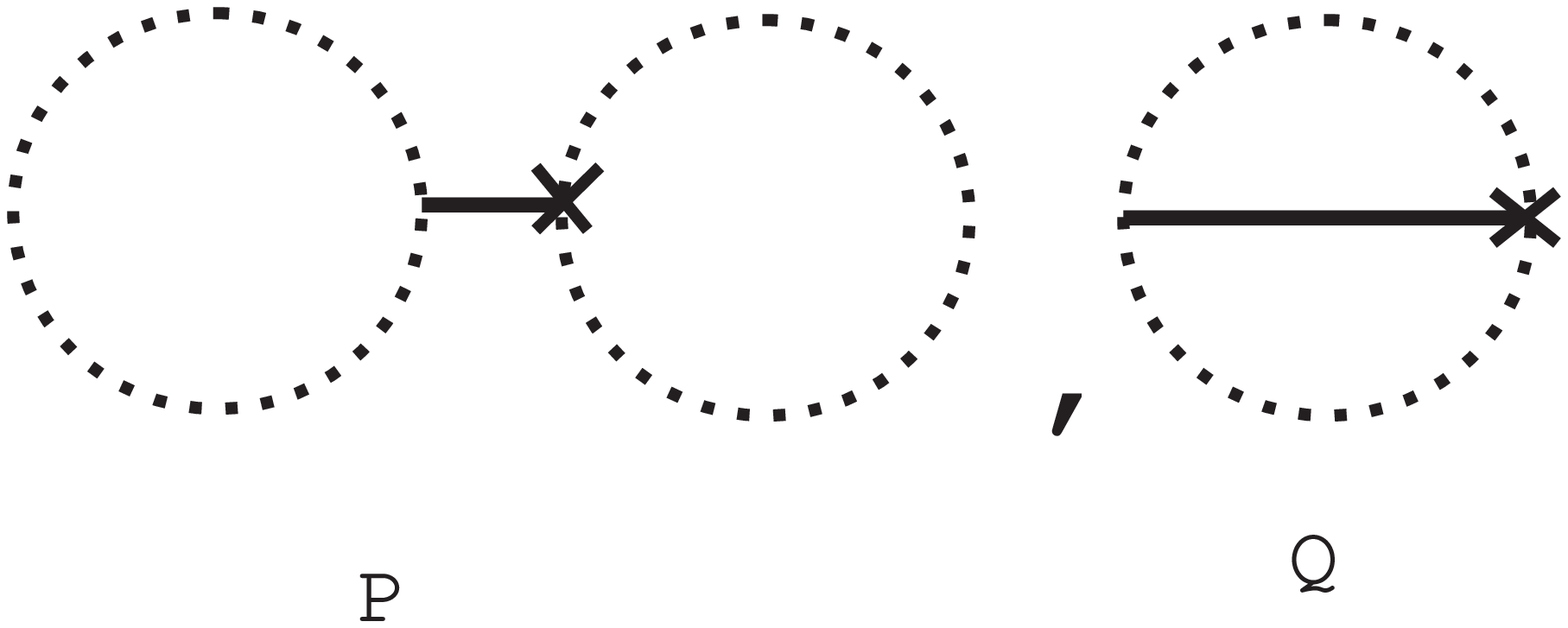} 
                 }
\caption{
Invariants of 
$P\equiv \pl_\m\pl_\m h_{\al\al}$\ and $Q\equiv \pl_\al\pl_\be h_{\ab}\ $.
        }
\label{fig3}
\end{figure}
%%%%%%%%%%%%%%%%%%%%%%%%%%%%%%%%%%%%%%%%%%%%%%%%%%%%%%%%%%%%%%%%%%%%%
Two independent  invariants (0-tensors)\ :\ 
$P\equiv \pl_\m\pl_\m h_{\al\al},\ Q\equiv \pl_\al\pl_\be h_{\ab}\ $
, which are made from Fig.2 by connecting the remaining two suffix-lines, are
expressed as in Fig.3.
They are all possible invariants  of $\pl\pl h$-type.
Generally all suffix-lines of invariants are  closed.
Those closed suffix-lines are called {\it suffix-loops}.

As for those tensors or invariants
which are made from two or more $\pl\pl h$-tensors, 
their graphical expressions are obtained in the same way. 
(See the following sections.) 
We see a tensor and
the corresponding graph are one-to-one.

%%%%%%%%%%%%%%%%%%%%%%%%%%%%%%%%%%%%%%%%%%%%%%%%%%%%%%%%%%%%%%%%%%%%%
%%%%%%%%%%%%%%%%%%%%%%%%%%%%%%   SEC  3    %%%%%%%%%%%%%%%%%%%%%%%%%%
%%%%%%%%%%%%%%%%%%%%%%%%%%%%%%%%%%%%%%%%%%%%%%%%%%%%%%%%%%%%%%%%%%%%%
\section{Code for Graphs}
%%%%%%%%%%%%%%%%%%%%  SEC 3.1  ddh-tensors  %%%%%%%%%%%%%%%%%%%%%%%%%
\subsection{Code for $\pl\pl h$-Tensors}
The basic element is the 4-tensor\ :
$\pl_1\pl_2 h_{34}$. 
Here and in the following we change the Greek symbols, $\m,\n,\al,\be,$
etc., to the non-negative integer numbers, $0,1, 2,3,4,$etc., 
for the convenience of programming.
%%%%%%%%%%%%%%%%%%%%%%%% Fig.4 %%%%%%%%%%%%%%%%%%%%%%%%%%%%%%%%%
\begin{figure}
     \centerline{
\psbox[width=95mm,height=50mm]{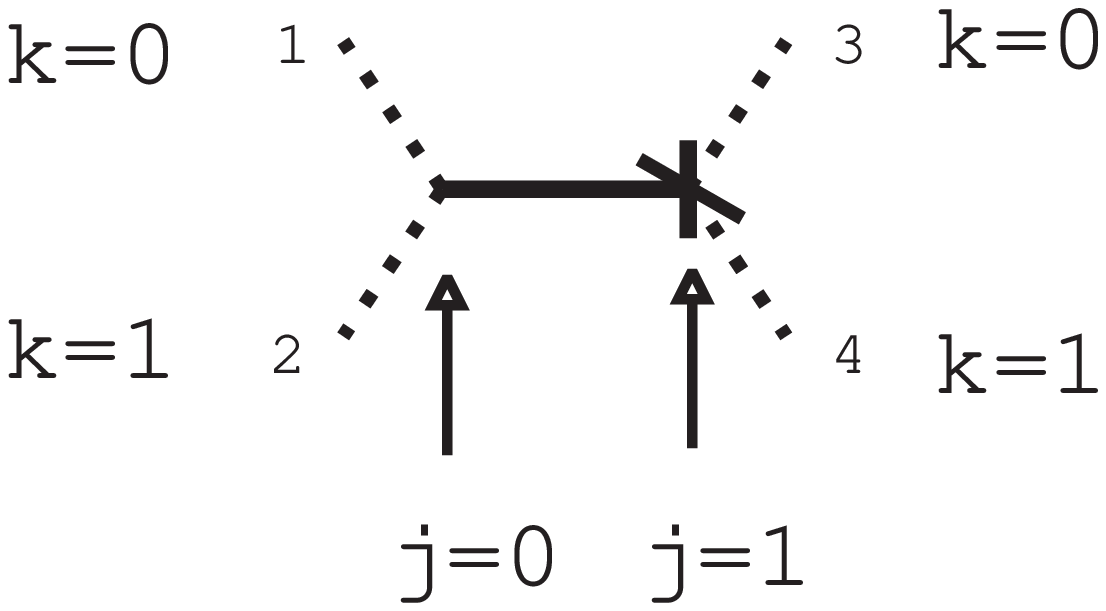}
                 }
\caption{
4-tensor of $\pl_1\pl_2 h_{34}$. 
        }
\label{fig4}
\end{figure}
%%%%%%%%%%%%%%%%%%%%%%%%%%%%%%%%%%%%%%%%%%%%%%%%%%%%%%%%%%%%%%%%%%%%%
The four suffixes, characterizing this tensor, are stored in a 3 dimensional
array of size $10\times 2\times 2$\ :\ d2h[i][j][k] as
%%%%%%%%%%%%%%%%%%%%%% Code for ddh %%%%%%%%%%%%%%%%%%%%%%%%%%%%%%%
\begin{tabbing}
d2h[i][0][0] \= = \= 1 \\
d2h[i][0][1] \> = \> 2 \\
d2h[i][1][0] \> = \> 3 \\
d2h[i][1][1] \> = \> 4 
\end{tabbing}
%%%%%%%%%%%%%%%%%%%%%%%%%%%%%%%%%%%%%%%%%%%%%%%%%%%%%%%%%%%%%%%%%%%%%%
where we regard this tensor as
``i''-th 'bond' ($\pl\pl h$-tensor), 
``j'' specifies a chosen vertex-type 
(j=0 for the vertex without $\times$, j=1 for the one with $\times$) 
of the i-th $\pl\pl h$-tensor and
``k'' specifies one of two suffix-lines at each vertex (i,j).
\footnote{
There exists arbitrariness in the choice of the input data due to
the suffix-permutation symmetry of the tensor $\pl_\m\pl_\n h_{\ab}$.
Because the graph identification is programmed 
in terms of its topology, we may take 
any one, say, 
\begin{tabbing}
d2h[i][0][0] \= = \= 2 \\
d2h[i][0][1] \> = \> 1 \\
d2h[i][1][0] \> = \> 4 \\
d2h[i][1][1] \> = \> 3. 
\end{tabbing}
         }
\footnote{
As the size of "i", we here take 10. This allows us to treat up to
10 different $\pl\pl h$-tensors at once. 
}
See Fig.4. We call "i","j" and "k" as {\it bond number}, 
{\it vertex-type number} and {\it suffix number} respectively.
%%%%%%%%%%%%%%%%%%%%%%%% Fig.5 %%%%%%%%%%%%%%%%%%%%%%%%%%%%%%%%%
\begin{figure}
     \centerline{
\psbox[width=50mm,height=45mm]{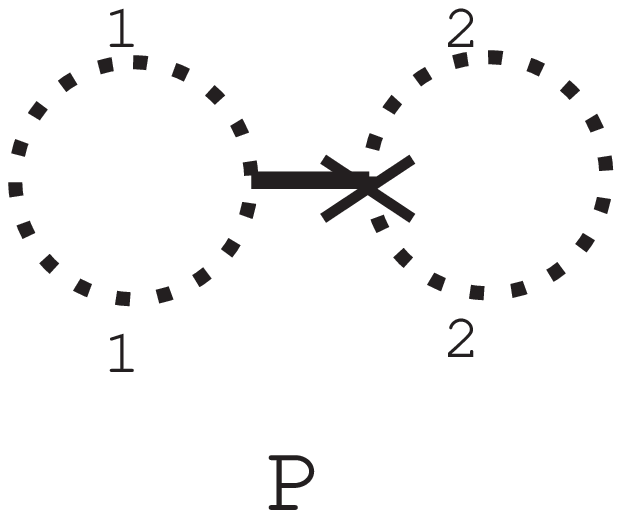}\q 
\psbox[width=28mm,height=40mm]{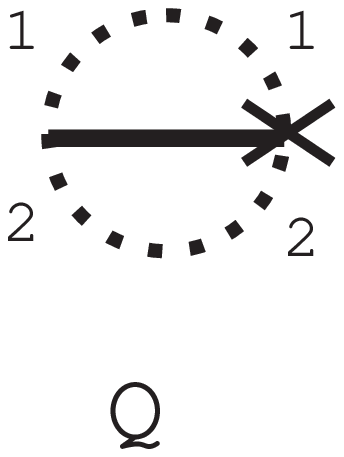} 
                 }
\caption{
Invariants of $P=\pl_1\pl_1 h_{22}$~and $Q=\pl_1\pl_2 h_{12}$\ . 
        }
\label{fig5.eps}
\end{figure}
%%%%%%%%%%%%%%%%%%%%%%%%%%%%%%%%%%%%%%%%%%%%%%%%%%%%%%%%%%%%%%%%%%%%%

In the examples of Fig.5, we store
two invariants, $P=\pl_1\pl_1 h_{22}$\ and $Q=\pl_1\pl_2 h_{12}$\  
in i=0 and i=1 component of d2h[i][\ ][\ ] respectively.
%%%%%%%%%%%%%%%%%%%%%% Code for P , Q %%%%%%%%%%%%%%%%%%%%%%%%%%%%%%%
\begin{tabbing}
d2h[0][0][0] \= = \= 1 \= \hspace{6.5cm} \= d2h[1][0][0] \= = \= 1 \\
d2h[0][0][1] \> = \> 1 \> \hspace{6.5cm} \> d2h[1][0][1] \> = \> 2 \\
d2h[0][1][0] \> = \> 2 \> \hspace{6.5cm} \> d2h[1][1][0] \> = \> 1 \\
d2h[0][1][1] \> = \> 2 \> \hspace{6.5cm} \> d2h[1][1][1] \> = \> 2 
\end{tabbing}
%%%%%%%%%%%%%%%%%%%%%%%%%%%%%%%%%%%%%%%%%%%%%%%%%%%%%%%%%%%%%%%%%%%%%%
Generally, contracted suffixes are {\it dummy} ones, that is, any suffix-numbers
may be taken unless the same number is used for different contractions.
In both examples above, the suffixes 1 and 2 are dummy ones and they can be
substituted by any non-negative integers, say, 3 and 4 respectively. 

%%%%%%%%%%%%%%%%%%%%  SEC 3.2  (ddh)*(ddh)-tensors  %%%%%%%%%%%%%%%%%%
\subsection{Code for $(\pl\pl h)^2$-Tensors}

%%%%%%%%%%%%%%%%%%%%%%%% Fig.6 %%%%%%%%%%%%%%%%%%%%%%%%%%%%%%%%%
\begin{figure}
     \centerline{
\psbox[width=110mm,height=40mm]{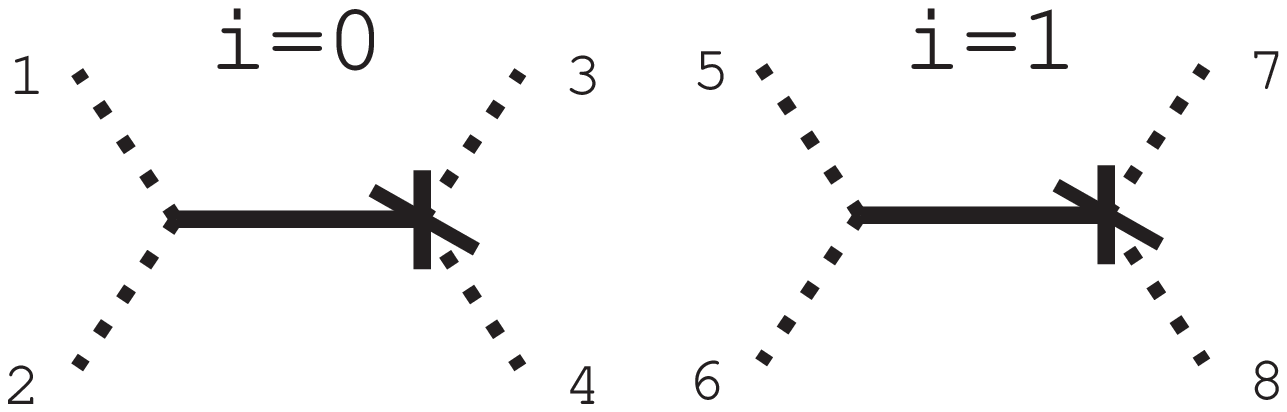}
                 }
\caption{
8-tensor of $\pl_1\pl_2 h_{34}\cdot\pl_5\pl_6 h_{78}$.
        }
\label{fig6}
\end{figure}
%%%%%%%%%%%%%%%%%%%%%%%%%%%%%%%%%%%%%%%%%%%%%%%%%%%%%%%%%%%%%%%%%%%%%
The most general $(\pl\pl h)^2$-type tensor is the 8-tensor:\ 
$\pl_1\pl_2 h_{34}\cdot\pl_5\pl_6 h_{78}$ which is graphically
shown in Fig.6.
The code for Fig.6 is
\footnote{
Because the numbering of two bonds is arbitrary ( permutation
symmetry of two $\pl\pl h$-tensors in a $(\pl\pl h)^2$-tensor ),
we may take, say,
\begin{tabbing}
d2h[0][0][0] \= = \= 5 \= \qqqq    \= d2h[1][0][0] \= = \= 1 \\
d2h[0][0][1] \> = \> 6 \>          \> d2h[1][0][1] \> = \> 2 \\
d2h[0][1][0] \> = \> 7 \>          \> d2h[1][1][0] \> = \> 3 \\
d2h[0][1][1] \> = \> 8 \>          \> d2h[1][1][1] \> = \> 4
\end{tabbing}
}
%%%%%%%%%%%%%%%%%%%%%% Code for ddh*ddh %%%%%%%%%%%%%%%%%%%%%%%%%%%%%%%
\begin{tabbing}
d2h[0][0][0] \= = \= 1 \= \qqqq    \= d2h[1][0][0] \= = \= 5 \\
d2h[0][0][1] \> = \> 2 \>          \> d2h[1][0][1] \> = \> 6 \\
d2h[0][1][0] \> = \> 3 \>          \> d2h[1][1][0] \> = \> 7 \\
d2h[0][1][1] \> = \> 4 \>          \> d2h[1][1][1] \> = \> 8 
\end{tabbing}
%%%%%%%%%%%%%%%%%%%%%%%%%%%%%%%%%%%%%%%%%%%%%%%%%%%%%%%%%%%%%%%%%%%%%%

A 4-tensor $\pl_1\pl_2 h_{34}\cdot\pl_5\pl_6 h_{34}$, which is made from
the above 8-tensor(Fig.6) by contracting(connecting) the suffixes
(suffix-lines) 3 and 4 with 7 and 8 respectively, is represented as in Fig.7.
%%%%%%%%%%%%%%%%%%%%%%%% Fig.7 %%%%%%%%%%%%%%%%%%%%%%%%%%%%%%%%%
\begin{figure}
     \centerline{
\psbox[width=65mm,height=35mm]{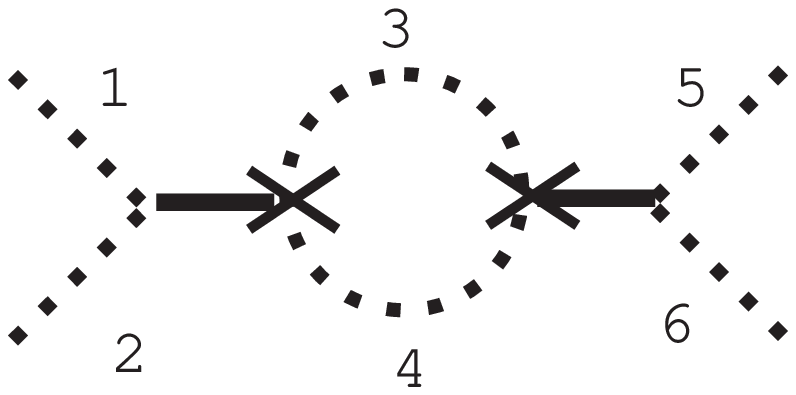}
                 }
\caption{
4-tensor of~$\pl_1\pl_2 h_{34}\cdot\pl_5\pl_6 h_{34}$. 
        }
\label{fig7}
\end{figure}
%%%%%%%%%%%%%%%%%%%%%%%%%%%%%%%%%%%%%%%%%%%%%%%%%%%%%%%%%%%%%%%%%%%%%
The code for Fig.7 is
%%%%%%%%%%%%%%%%%%%%%% Code for ddh*ddh %%%%%%%%%%%%%%%%%%%%%%%%%%%%%%%
\begin{tabbing}
d2h[0][0][0] \= = \= 1 \= \qqqq \= d2h[1][0][0] \= = \= 5 \\
d2h[0][0][1] \> = \> 2 \>       \> d2h[1][0][1] \> = \> 6 \\
d2h[0][1][0] \> = \> 3 \>       \> d2h[1][1][0] \> = \> 3 \\
d2h[0][1][1] \> = \> 4 \>       \> d2h[1][1][1] \> = \> 4
\end{tabbing}
%%%%%%%%%%%%%%%%%%%%%%%%%%%%%%%%%%%%%%%%%%%%%%%%%%%%%%%%%%%%%%%%%%%%%%
where two indices 3 and 4 are dummy. There are some other 4-tensors
which can be obtained from Fig.6 by connecting two pairs of suffix-lines
in  different ways. 

An invariant:\ $\pl_1\pl_2 h_{34}\cdot\pl_1\pl_2 h_{34}$, which is made from
Fig.7 by contracting suffixes 1 and 2 with 5 and 6 respectively, is
graphically obtained as in Fig.8.
%%%%%%%%%%%%%%%%%%%%%%%% Fig.8 %%%%%%%%%%%%%%%%%%%%%%%%%%%%%%%%%
\begin{figure}
     \centerline{
\psbox[width=62mm,height=37mm]{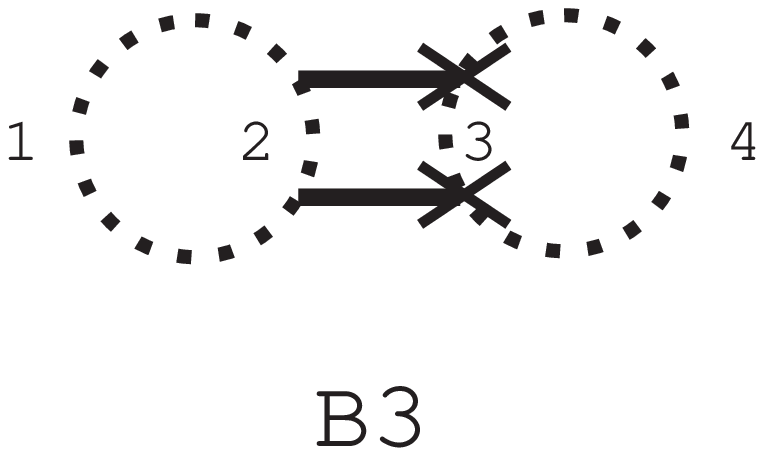}
                 }
\caption{
Invariant of $B3\equiv\pl_1\pl_2 h_{34}\cdot\pl_1\pl_2 h_{34}$.
        }
\label{fig8}
\end{figure}
%%%%%%%%%%%%%%%%%%%%%%%%%%%%%%%%%%%%%%%%%%%%%%%%%%%%%%%%%%%%%%%%%%%%%
The code for Fig.8 is
%%%%%%%%%%%%%%%%%%%%%% Code for ddh*ddh %%%%%%%%%%%%%%%%%%%%%%%%%%%%%%%
\begin{tabbing}
d2h[0][0][0] \= = \= 1 \= \qqqq \= d2h[1][0][0] \= = \= 2 \\
d2h[0][0][1] \> = \> 2 \>       \> d2h[1][0][1] \> = \> 1 \\
d2h[0][1][0] \> = \> 4 \>       \> d2h[1][1][0] \> = \> 3 \\
d2h[0][1][1] \> = \> 3 \>       \> d2h[1][1][1] \> = \> 4
\end{tabbing}
%%%%%%%%%%%%%%%%%%%%%%%%%%%%%%%%%%%%%%%%%%%%%%%%%%%%%%%%%%%%%%%%%%%%%%

All possible and independent invariants of $(\pl\pl h)^2$-type are listed
in Appendix A. 
As mentioned before, all the suffix-lines of
the invariants are closed.

%%%%%%%%%%%%%%%%%%%%%%%%%%%%%%%%%%%%%%%%%%%%%%%%%%%%%%%%%%%%%%%%%%%
%%%%%%%%%%%%%%%%%%%%%%%%%%%%%   SEC  4  %%%%%%%%%%%%%%%%%%%%%%%%%%%
%%%%%%%%%%%%%%%%%%%%%%%%%%%%%%%%%%%%%%%%%%%%%%%%%%%%%%%%%%%%%%%%%%%
\section{Indices for Characterizing $(\pl\pl h)^2$-Invariants}
%%%%%%%%%%%%%%%%%%%%%%%%%%%%%%%%%%%%%%%%%%%%%%%%%%%%%%%%%%%%%%%%%%%%%
By virtue of the graphical representation, we can identify a tensor
by the topology of the corresponding graph. 
We need not to know the explicit suffix-names associated with the
tensor any more. In the computer calculation, 
this approach reduces the memory space and speeds up the tensor
calculation considerably.
Let us explain how to do it concretely using
the codes explained in the previous section.

The following two items are the important steps for the tensor calculation.
%%%%%%%%%%%%%%%%%%% Two items   %%%%%%%%%%%%%%%%%%%%%%
\begin{itemize}
\item
Listing up all independent invariants (closed graphs).
\item
Characterizing every closed graph by a set of some 
appropriate {\it indices}, 
which are computed from the code of the graph. 
Those indices must
be uniquely defined irrespective of arbitrariness of the code due to \ 
1)\ the suffix permutation symmetry,\  
2)\ arbitrariness in numbering bonds, suffix-loops, 
{\it tadpoles}(see later), etc,\ 
3)\ arbitrariness of dummy suffixes,\ 
4)\ others (see later).  
\end{itemize}
%%%%%%%%%%%%%%%%%%%%%%%%%%%%%%%%%%%%%%%%%%%%%%%%%%%%%%%%
The former
is generally done by the graph theory \cite{Harary}, while the latter 
can be done by a 
program using the present algorithm. The {\it indices} are explained
in this section.

Let us begin with the simplest case of  $\pl\pl h$-invariants.
See Fig.5.
We denote, [the number of suffix-loops] $-$1, as 
\ul{loop} in the program.
\footnote{
In the following, when the variable names, used in the program, 
appear in ordinary sentences of the text, they are underlined
('\ul{\q\q}').
}
The index of \ul{loop}  is sufficient for discriminating two independents,
P and Q;\ 
\ul{loop}=1 for $P=\pl_1\pl_1 h_{22}$\ and 
\ul{loop}=0 for $Q=\pl_1\pl_2 h_{12}$.
In this paper 
we focus only on $(\pl\pl h)^2$-invariants
in order to explain the algorithm most simply.
The generalization to the general $(\pl\pl h)^m$-invariants
 (m=1,2,3,$\cdots$) is straightforward. 
(See Sec.6.) 
For the $(\pl\pl h)^2$-invariants,
we can easily check, without the knowledge of the graph theory, 
that 13 independent invariants listed in App.A exhaust 
all possible independent ones\cite{II,II2}. 
Because all suffix-lines for  invariants are always closed, 
we seek some good indices 
associated with each closed loop.
Of course the number of closed loops, \ul{loop}+1,  
is one good index. It is, however, 
not sufficient to discriminate 13 invariants completely.

%%%%%%%%%%%%%%%%%%%%%  4.1  %%%%%%%%%%%%%%%%%%%%%%%%%%%%%%%%%%%%%%%%%
\subsection{\ul{loop}, \ul{loopstream}[\ ][\ ][\ ] and \ul{loopno}[\ ][\ ]}
A 3-dim array 
\ul{loopstream}[\ul{maxloopno}][2][2], defined below, is initialized as

%%%%%%%%%%%%%%%%%%%%%% Initialization of loopstream %%%%%%%%%%%%%%
\begin{quote}
\centering
loopstream[$l$][i][j]  = 99, \\
$l$=0,1,$\cdots\ <$maxloopno$\equiv$4;\ i=0,1;\ j=0,1 \\
\end{quote}
%%%%%%%%%%%%%%%%%%%%%%%%%%%%%%%%%%%%%%%%%%%%%%%%%%%%%%%%%%%%%%%%%%%%%%
For $l$-th loop, we assign an ascending natural number, 1,2,3,$\cdots$, at
each vertex (i,j) along the loop:\ \ul{loopstream}[$l$][i][j]=1,2,3,$\cdots$.
\footnote{
The choice of the starting vertex is arbitrary. This is one of
``4) others'' in the list of arbitrariness, in the beginning of
this section.
The number "i" refers to "i-th bond" and it runs from 0 to 1 because
we are considering a quadratic term of $\pl\pl h$. The number "j" refers
to a vertex-type: j=0 means the vertex without the crossing symbol, j=1 means
the vertex with it. See Fig.1.
}
When a vertex (i,j) belongs to $l$-th loop, we assign as 
\ul{loopno}[i][j]=$l$. See Fig.9 for example.

%%%%%%%%%%%%%%%%%%%%%%%% Fig.9 %%%%%%%%%%%%%%%%%%%%%%%%%%%%%%%%%
\begin{figure}
     \centerline{
\psbox[width=145mm,height=45mm]{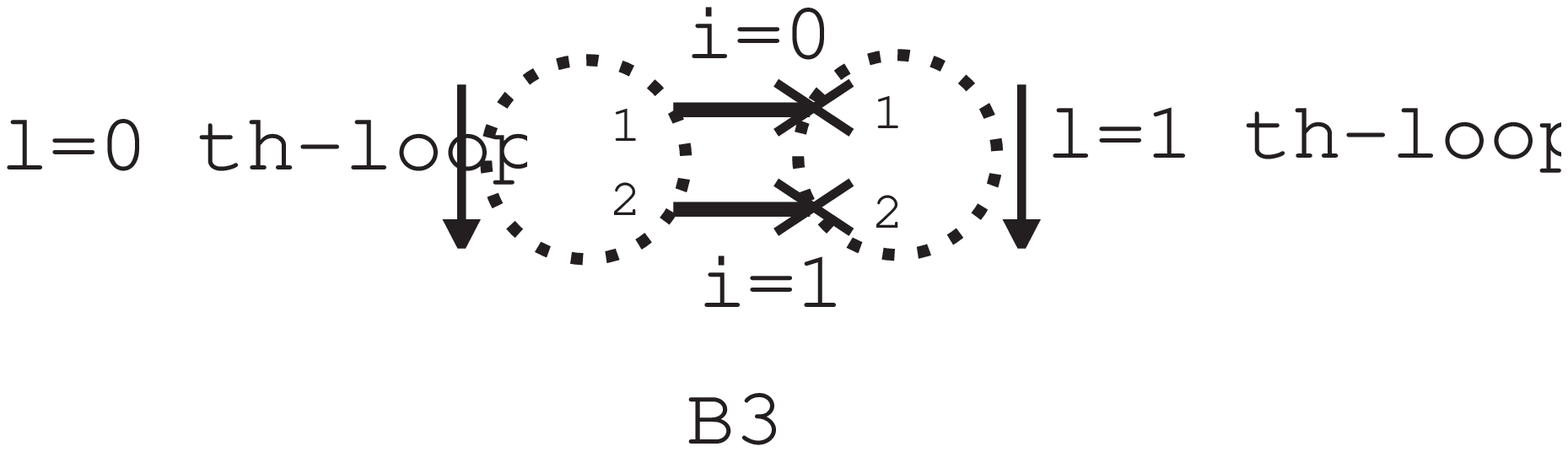}
                 }
\caption{
\ul{loopstream}[\ ][\ ][\ ] for $B3$. 
Arrows indicate the directions of tracing along each loop.
        }
\label{fig9}
\end{figure}
%%%%%%%%%%%%%%%%%%%%%%%%%%%%%%%%%%%%%%%%%%%%%%%%%%%%%%%%%%%%%%%%%%%%%

%%%%%%%%%%%%%%%%%%%%%% loopstream[][][],loopno[][] %%%%%%%%%%%%%%%%%%
\begin{tabbing}
loop=1 \\
loopstream[0][0][0] \= = \= 1  \= \qqq \= loopno[0][0]=0   \\
loopstream[0][0][1] \> = \> 99 \> \qqq \> loopno[0][1]=1   \\
loopstream[0][1][0] \> = \> 2  \> \qqq \> loopno[1][0]=0   \\
loopstream[0][1][1] \> = \> 99 \> \qqq \> loopno[1][1]=1   \\
loopstream[1][0][0] \> = \> 99 \> \> \\
loopstream[1][0][1] \> = \> 1  \> \> \\
loopstream[1][1][0] \> = \> 99 \> \> \\
loopstream[1][1][1] \> = \> 2  \> \>
\end{tabbing}
%%%%%%%%%%%%%%%%%%%%%%%%%%%%%%%%%%%%%%%%%%%%%%%%%%%%%%%%%%%%%%%%%%%%%%
%%%%%%%%%%%%%%%%%%%%%  4.2  %%%%%%%%%%%%%%%%%%%%%%%%%%%%%%%%%%%%%%%%%
\subsection{\ul{maxver}[\ ], \ul{vertex}[\ ][\ ][\ ] }
\ul{maxver}[$l$] stores the number of vertices which $l$-th loop has.
When \ul{maxver}[$l'$]=1, the $l'$-th loop is called {\it tadpole}. See
Fig.10. 
%%%%%%%%%%%%%%%%%%%%%%%% Fig.10 %%%%%%%%%%%%%%%%%%%%%%%%%%%%%%%%%
\begin{figure}
     \centerline{
\psbox[width=85mm,height=40mm]{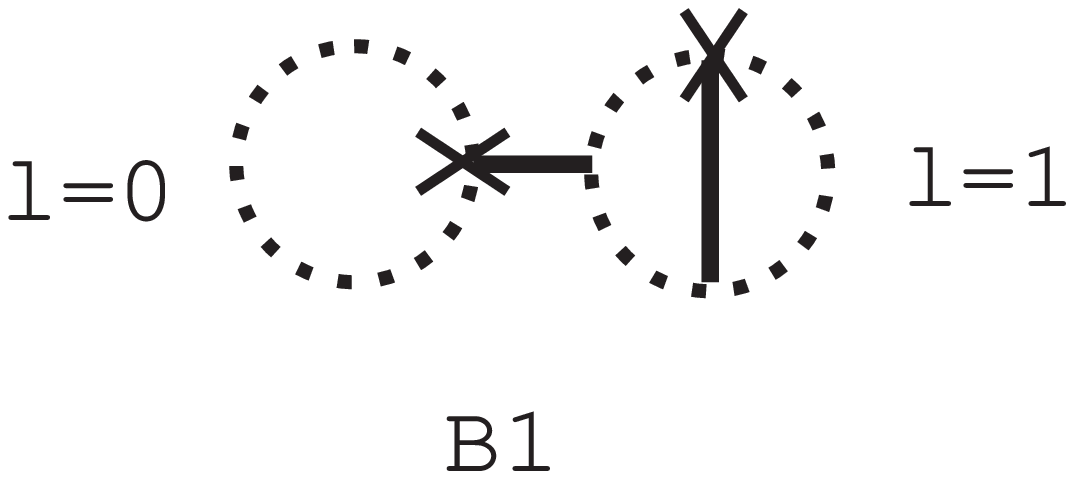}
                 }
\caption{
Invariant of $B1$. \# of vertices of 0-th and 1-th loop
are 1 and 3 respectively. 0-th loop is a tadpole.
        }
\label{fig10}
\end{figure}
%%%%%%%%%%%%%%%%%%%%%%%%%%%%%%%%%%%%%%%%%%%%%%%%%%%%%%%%%%%%%%%%%%%%%
\ul{maxver}[\ ] for Fig.10 is given by
%%%%%%%%%%%%%%%%%%%%%% maxver[] %%%%%%%%%%%%%%
\begin{quote}
%\centering
\raggedright
maxver[0]  = 1 \\
maxver[1]  = 3 \\
\end{quote}
%%%%%%%%%%%%%%%%%%%%%%%%%%%%%%%%%%%%%%%%%%%%%%%%%%%%%%%%%%%%%%%%%%%%%%
For later convenience, 
the same information as \ul{loopstream}[\ ][\ ][\ ] is stored in another
form:\ 
\ul{vertex}[\ ][\ ][\ ]. When we trace the l-th loop in a certain direction
\footnote{
The choice of the direction of tracing is another arbitrariness in 
``4) others'' in the beginning of this section.
}
,
let the k-th vertex (k=0,1,$\cdots$) be (i,j). Then we assign as
\ul{vertex}[$l$][k][0]=i, \ul{vertex}[$l$][k][1]=j,\ 
k=0,1,$\cdots <$ maxver[$l$]. For Fig.9
we have
%%%%%%%%%%%%%%%%%%%%%% vertex[][][] %%%%%%%%%%%%%%
\begin{quote}
\centering
vertex[0][0][0]  = 0 \\
vertex[0][0][1]  = 0 \\
vertex[0][1][0]  = 1 \\
vertex[0][1][1]  = 0 \\
vertex[1][0][0]  = 0 \\
vertex[1][0][1]  = 1 \\
vertex[1][1][0]  = 1 \\
vertex[1][1][1]  = 1 \\
\end{quote}
%%%%%%%%%%%%%%%%%%%%%%%%%%%%%%%%%%%%%%%%%%%%%%%%%%%%%%%%%%%%%%%%%%%%%%
%%%%%%%%%%%%%%%%%%%%  4.3  %%%%%%%%%%%%%%%%%%%%%%%%%%%%%%%%%%%%%%%%%
\subsection{\ul{absdeli}[\ ], \ul{absdelj}[\ ], \ul{tadpoleno} 
and \ul{tad}[\ ][\ ]}
%%%%%%%%%%%%%%%%%%%%%%%% Fig.11 %%%%%%%%%%%%%%%%%%%%%%%%%%%%%%%%%
\begin{figure}
     \centerline{
\psbox[width=75mm,height=45mm]{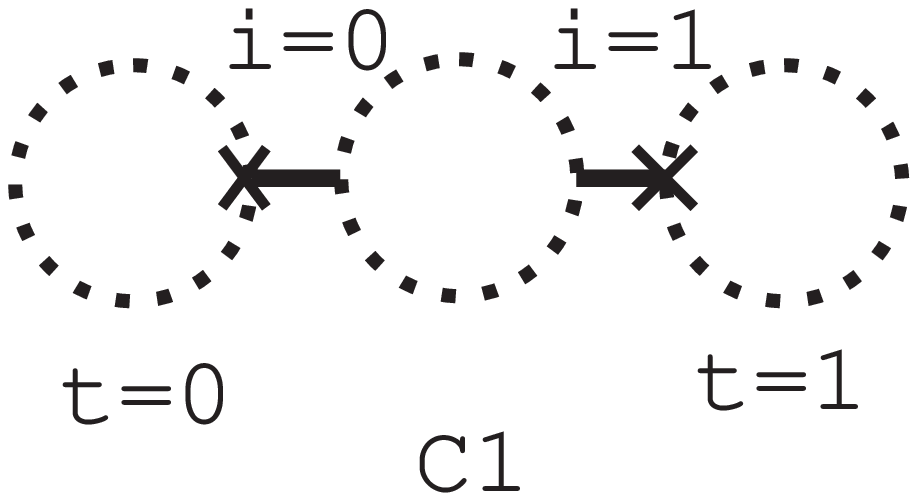}\ \ 
\psbox[width=75mm,height=45mm]{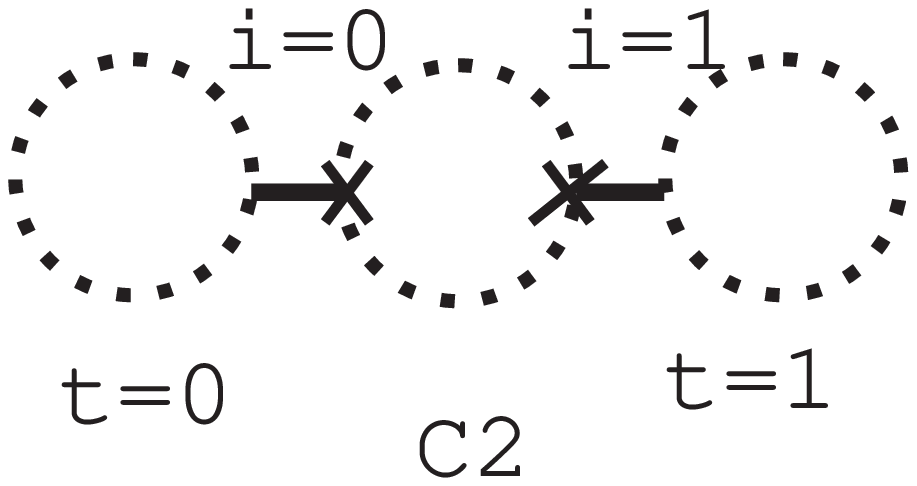}
                }
\vspace{2mm}
     \centerline{				 
\psbox[width=75mm,height=45mm]{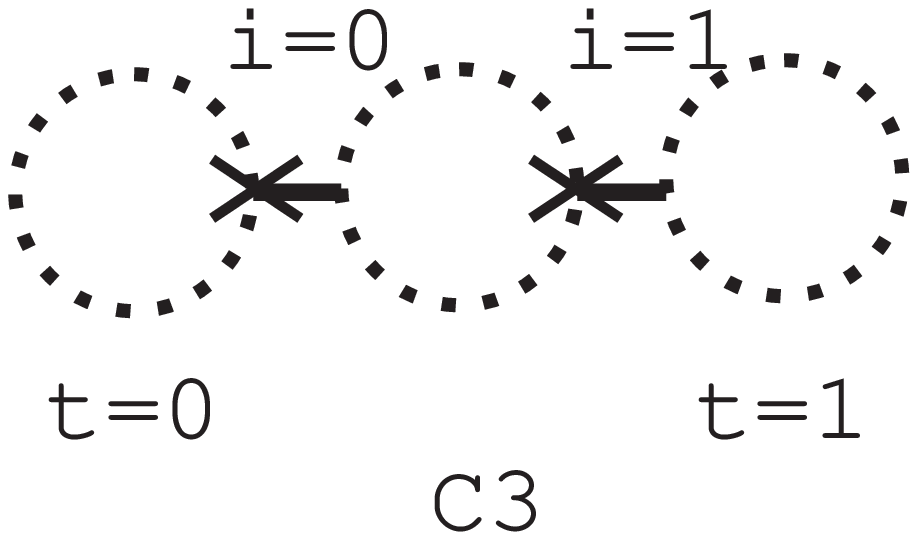} 
                 }
\caption{
\ul{tadpoleno} and \ul{tad}[\ ][\ ] for $C1, C2$~and $C3$. 
        }
\label{fig11}
\end{figure}
%%%%%%%%%%%%%%%%%%%%%%%%%%%%%%%%%%%%%%%%%%%%%%%%%%%%%%%%%%%%%%%%%%%%%

The number of tadpole-loops in a graph is 
stored in \ul{tadpoleno}. When the
t-th\lb (t=0,1,$\cdots <$\ul{tadpoleno}) tadpole-loop connects with a bond
at a vertex (i,j), we assign \ul{tad}[t][0]=i, \ul{tad}[t][1]=j.
For C1, C2 and C3 of Fig.11, we have the following. 
%%%%%%%%%%%%%%%%%%%%%% tadpoleno, tad[][] %%%%%%%%%%%%%%%%%%%%%%%%%%%%%%%
\begin{tabbing}
\q\q\q\q  \= \q C1\q \= C2\q  \= C3   \\
tadpoleno \> \q 2\q  \> 2\q   \> 2   \\
tad[0][0] \> \q 0\q  \> 0\q   \> 0   \\
tad[0][1] \> \q 1\q  \> 0\q   \> 1   \\
tad[1][0] \> \q 1\q  \> 1\q   \> 1   \\
tad[1][1] \> \q 1\q  \> 0\q   \> 0   
\end{tabbing}
%%%%%%%%%%%%%%%%%%%%%%%%%%%%%%%%%%%%%%%%%%%%%%%%%%%%%%%%%%%%%%%%%%%%%%

Now we come to  key quantities in this algorithm.
When we trace a suffix-line, along a loop, starting from a vertex 
(i$_0$,j$_0$) in a certain direction, we pass some vertices,
(i$_1$,j$_1$),(i$_2$,j$_2$),$\cdots$ and finally come back to the
starting vertex. We focus on the {\it change} of the bond number(i) and the
vertex-type number(j) when we pass from a vertex to the next vertex in 
the above tracing. See Fig.12.
%%%%%%%%%%%%%%%%%%%%%%%% Fig.12 %%%%%%%%%%%%%%%%%%%%%%%%%%%%%%%%%
\begin{figure}
     \centerline{
\psbox[width=135mm,height=145mm]{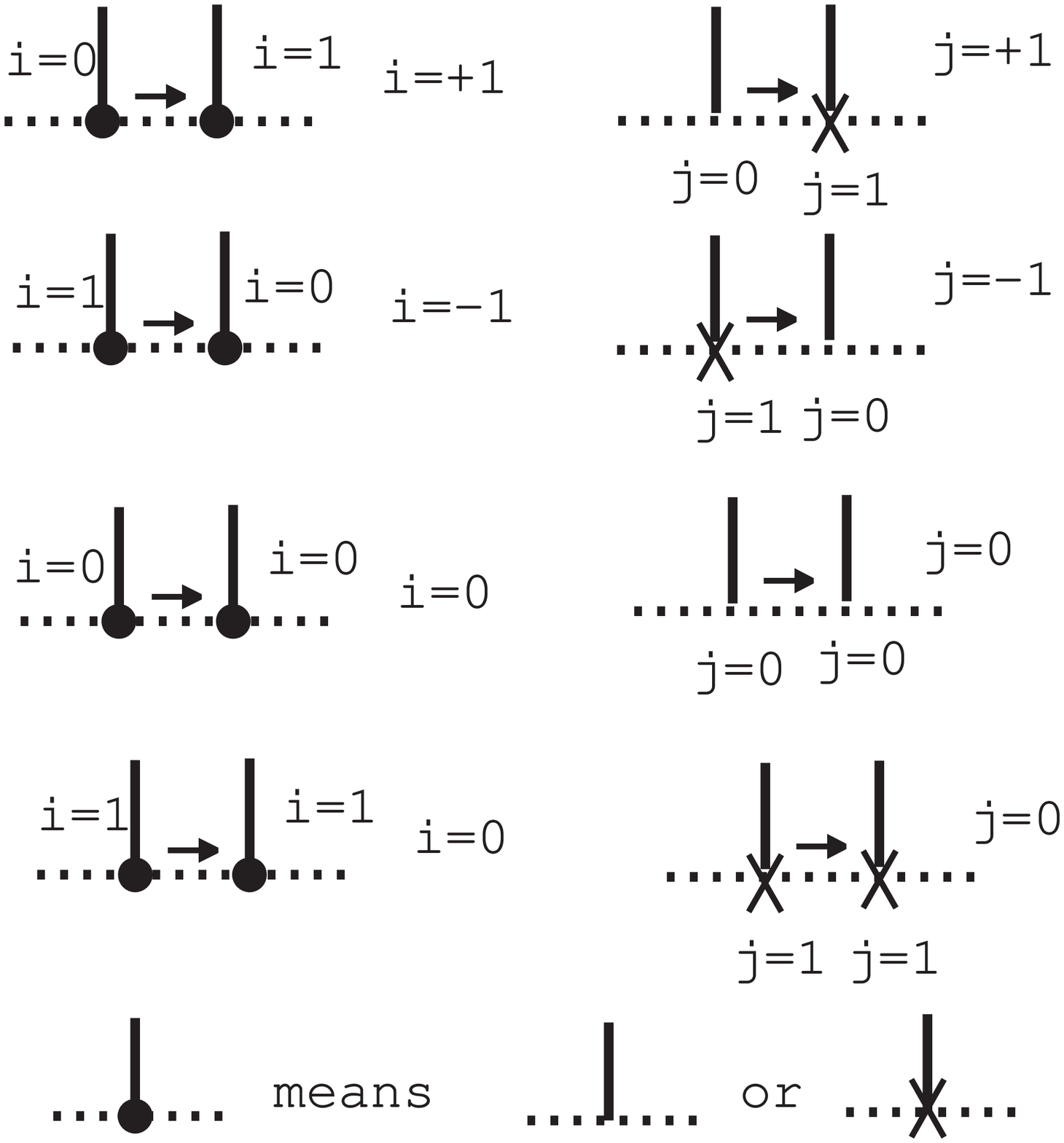}
                 }
\caption{
Change of i (bond number) and j (vertex-type number).\protect\nl
Arrows indicate directions of tracings.
        }
\label{fig12}
\end{figure}
%%%%%%%%%%%%%%%%%%%%%%%%%%%%%%%%%%%%%%%%%%%%%%%%%%%%%%%%%%%%%%%%%%%%%
For $l$-th loop, we assign as
$\sum_{\mbox{along $l$-loop}}|\Delta \mbox{i}|\equiv$ \ul{absdeli}[$l$],
$\sum_{\mbox{along $l$-loop}}|\Delta \mbox{j}|\equiv$ \ul{absdelj}[$l$].
(They are called {\it bond changing number} and {\it vertex changing number}
respectively, in Ref.\cite{II97}.)
As examples, the indices for A1, A2 and A3 of Fig.13 are given as
%%%%%%%%%%%%%%%%%%%%%% absdeli[],absdelj[] %%%%%%%%%%%%%%%%%%%%%%%%%%%%%%%
\begin{tabbing}
  \q\q\q\q \= \q A1\q      \= \q A2\q    \= \q A3   \\
\ul{loop}       \> \q 0\q       \> \q 0\q     \> \q  0   \\
\ul{absdeli}[0] \> \q 4\q       \> \q 2\q     \> \q  2   \\
\ul{absdelj}[0] \> \q 2\q       \> \q 2\q     \>  \q 4   
\end{tabbing}
%%%%%%%%%%%%%%%%%%%%%%%%%%%%%%%%%%%%%%%%%%%%%%%%%%%%%%%%%%%%%%%%%%%%%%

%%%%%%%%%%%%%%%%%%%%%%%% Fig.13 %%%%%%%%%%%%%%%%%%%%%%%%%%%%%%%%%
\begin{figure}
     \centerline{
\psbox[width=60mm,height=60mm]{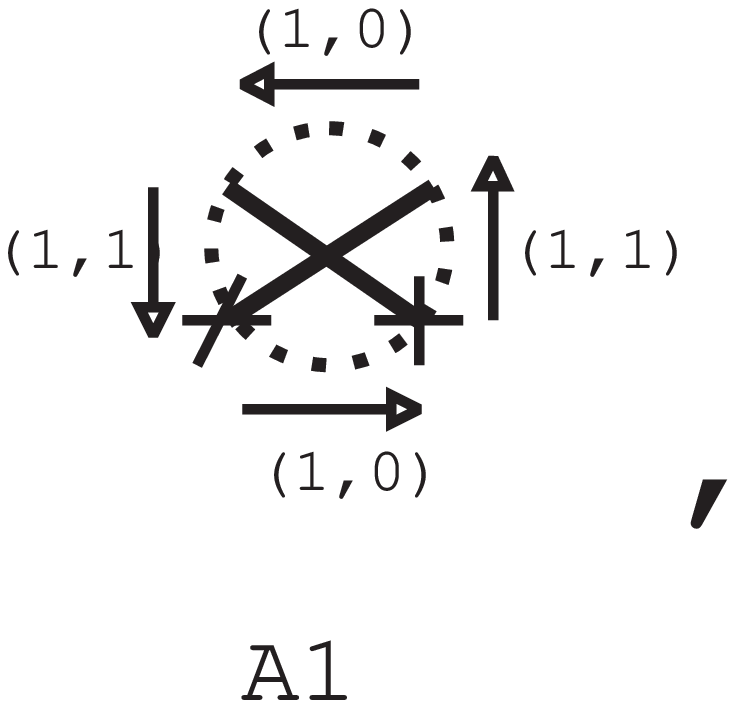}\ 
\psbox[width=55mm,height=60mm]{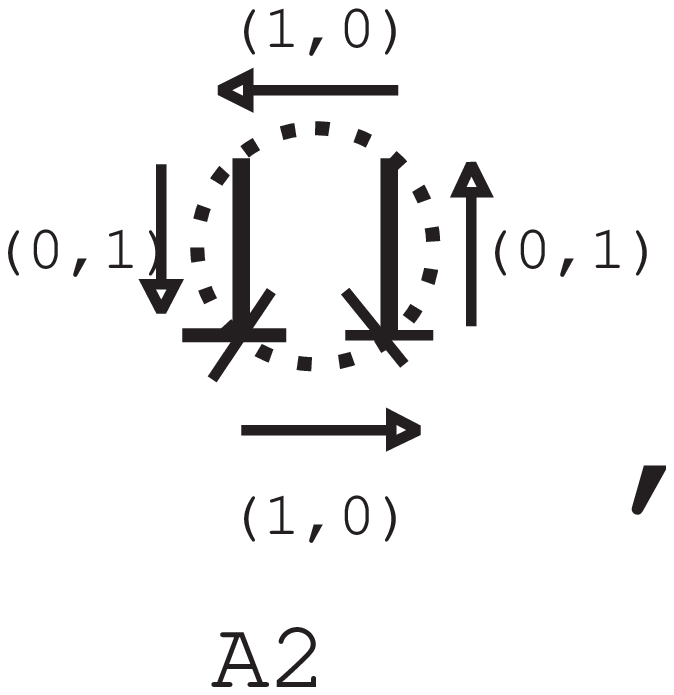}\ 
\psbox[width=55mm,height=60mm]{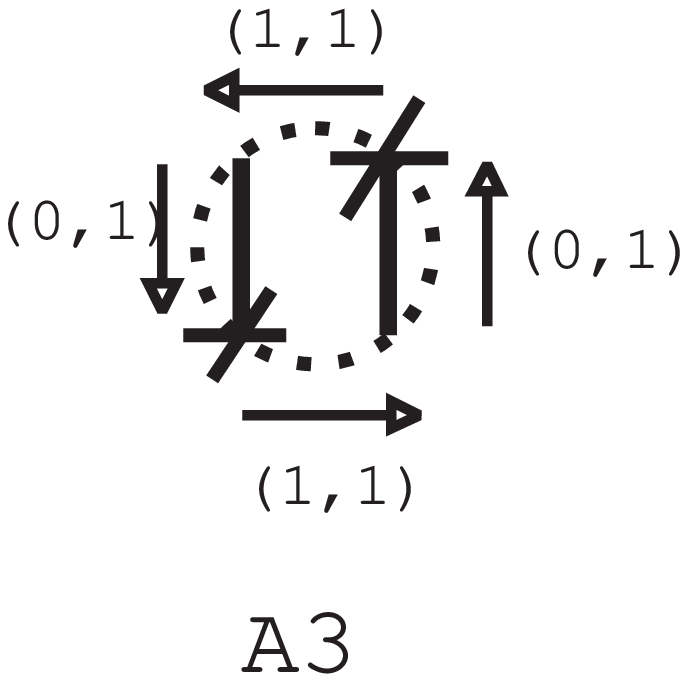} 
                 }
\caption{
Calculation of 
\ul{absdeli}[\ ] and \ul{absdelj}[\ ] for $A1, A2$~and $A3$.
($|\Del i|,|\Del j|$) are listed. 
        }
\label{fig13}
\end{figure}
%%%%%%%%%%%%%%%%%%%%%%%%%%%%%%%%%%%%%%%%%%%%%%%%%%%%%%%%%%%%%%%%%%%%%

\ul{absdeli}[\ ] and \ul{absdelj}[\ ] defined above satisfy the following
important properties for programming.
\begin{enumerate}
\item
They do not depend on the starting vertex for tracing along a loop.
\item
They do not depend on the direction of the tracing.
\end{enumerate}

%%%%%%%%%%%%%%%%%%%%%  4.4  %%%%%%%%%%%%%%%%%%%%%%%%%%%%%%%%%%%%%%%%%
\subsection{Table of Indices for all invariants}

We give the full list of indices for all invariants in Table 1.
The program, WEAKGRAV, computes all indices for each term,
and identify it with a  graph following the table.

\vspace{0.5cm}

%%%%%%%%%%%%%%%%%%%%%%%%%%%%%%%%%%%%%%%%%%%%%%%%%%%%%%%%%%%%%%%%%%%%%%%%%%
%%%%%%%%%%%%%%%%%%%%%  Table 1   %%%%%%%%%%%%%%%%%%%%%%%%%%%%%%%%%%%%%%%%%
%%%%%%%%%%%%%%%%%%%%%%%%%%%%%%%%%%%%%%%%%%%%%%%%%%%%%%%%%%%%%%%%%%%%%%%%%%
\begin{tabular}{|c||c|c|c|c|c|}
\hline
Graph,(Run.No)\ $\backslash$\ Indices
& \ul{loop}& \ul{tadpoleno}& \ul{tad}[\ ][1]&\ul{absdeli}[\ ]
                                                          &\ul{absdelj}[\ ]\\
\hline
        &       &            &              &              &              \\
$A1
=\pl_\si\pl_\la h_\mn\cdot\pl_\si\pl_\n h_{\m\la}$,(0) 
        &  0    & 0          & nothing      & 4            & 2            \\
        &       &            &              &              &              \\
\hline
        &       &            &              &              &              \\
$A2
=\pl_\si\pl_\la h_{\la\m}\cdot\pl_\si\pl_\n h_{\mn}$,(1)
        & 0    & 0          & nothing      & 2            & 2            \\
        &       &            &              &              &              \\
\hline
        &       &            &              &              &              \\
$A3
=\pl_\si\pl_\la h_{\la\m}\cdot\pl_\m\pl_\n h_{\n\si}$,(2)
        & 0    & 0          & nothing      & 2            & 4            \\
        &       &            &              &              &              \\
\hline
\hline
        &       &            &              &              &              \\
$B1
=\pl_\n\pl_\la h_{\si\si}\cdot\pl_\la\pl_\m h_{\mn}$,(3)
        & 1    & 1          & 1            & $/$          & $/$          \\
        &       &            &              &              &              \\
\hline
        &       &            &              &              &              \\
$B2
=\pl^2 h_{\la\n}\cdot\pl_\la\pl_\m h_{\mn}$,(4)
        & 1    & 1          & 0            & $/$          & $/$          \\
        &       &            &              &              &              \\
\hline
        &       &            &              &  2          &  0          \\
\cline{5-6}
$B3
=\pl_\m\pl_\n h_{\la\si}\cdot\pl_\m\pl_\n h_{\ls}$,(5)
        & 1    & 0          & nothing      &   2          &  0          \\
\cline{5-6}
        &       &            &              &              &              \\
\hline
        &       &            &              &   2         &  2          \\
\cline{5-6}
$B4
=\pl_\m\pl_\n h_{\la\si}\cdot\pl_\la\pl_\si h_{\mn}$,(6)
        & 1    & 0          & nothing      &    2         &  2          \\
\cline{5-6}
        &       &            &              &              &              \\
\hline
        &       &            &              &   0         &  2           \\
\cline{5-6}
$Q^2
=(\pl_\m\pl_\n h_{\mn})^2$,(7)
        & 1     & 0          & nothing      &   0         &  2           \\
\cline{5-6}
        &       &            &              &              &               \\
\hline
\hline
        &       &            & 1             &    0        & 0              \\
\cline{4-6}
$C1
=\pl_\m\pl_\n h_{\la\la}\cdot\pl_\m\pl_\n h_{\si\si}$,(8)
        & 2    & 2          &                & 2          &  0           \\
\cline{4-6}
        &       &            & 1             & 0          &   0             \\
\hline
        &       &            &  0            & 0          &   0             \\
\cline{4-6}
$C2
=\pl^2 h_{\mn}\cdot\pl^2 h_{\mn}$,(9)
        & 2    & 2          &                &   2         &  0           \\
\cline{4-6}
        &       &            &  0            &   0         &  0            \\
\hline
        &       &            &  1            &   0         &  0           \\
\cline{4-6}
$C3
=\pl_\m\pl_\n h_{\la\la}\cdot\pl^2 h_{\mn}$,(10)
        & 2   & 2          & 0          &   0         &  0          \\
\cline{4-6}
        &       &            &              &   2         &  2          \\
\hline
        &       &            &     1         &   0         &  0           \\
\cline{4-6}
$PQ
=\pl^2 h_{\la\la}\cdot\pl_\m\pl_\n h_{\mn}$,(11)
        & 2   & 2          & 0          &   0         &  0          \\
\cline{4-6}
        &       &            &              &   0         &  2           \\
\hline
\hline
        &       &            &              &              &              \\
$P^2
=(\pl^2 h_{\la\la})^2$,(12)
        & 3  &  $/$       & $/$          & $/$          & $/$          \\
        &       &            &              &              &              \\
\hline
\multicolumn{6}{c}{\q}                                                 \\
\multicolumn{6}{c}{Table 1\ \  List of indices for all 
$(\pl\pl h)^2$-invariants. 
The symbol '$/$' means }\\
\multicolumn{6}{c}{'need not be calculated'.
Graph names, $A1,A2,\cdots$~are defined in Appendix A. 
}\\
\end{tabular}
%%%%%%%%%%%%%%%%%%%%%%%%%  END  of  Table 1 %%%%%%%%%%%%%%%%%%%%%%%%%%%%%
\newpage
%\vs 1
Note here that the quantities in Table 1 do not depend on 
the arbitrariness cited in the beginning of this section. 
This is one of key points of the present algorithm.

%\vs 1

%%%%%%%%%%%%%%%%%%%%%%%%%%%%%%%%%%%%%%%%%%%%%%%%%%%%%%%%%%%%%%%%%%
%%%%%%%%%%%%%%%%%%%%%%%%%%   SEC 5  %%%%%%%%%%%%%%%%%%%%%%%%%%%%%%%
%%%%%%%%%%%%%%%%%%%%%%%%%%%%%%%%%%%%%%%%%%%%%%%%%%%%%%%%%%%%%%%%%%%
\section{How to use WEAKGRAV}

Let us explain how to use WEAKGRAV taking some simple examples.
The program reads all input data from a file (named 'indata.dat').
The input and output data for two examples below are provided 
in Appendix B.

\begin{flushleft}
Ex.1\ $R_{\la\mn\si}R^{\la\mn\si}$
\end{flushleft}

The weak-field expansion 
($g_\mn=\del_\mn+h_\mn, |h_\mn|\ll 1$) of Riemann tensor is given as
\footnote{
See \cite{II} for the present convention of gravitational quantities.
}
%The input data can be read from
%****(how.1)%%%%%%%%%%%%%%%%%%%%
\begin{eqnarray}
2R_{\la\mn\si} &=& 2R^{\la\mn\si}+O'(h^2) \nn\\
               &=& \pl_\m\pl_\n h_\ls-\pl_\la\pl_\n h_{\m\si}
         -\n\change\si +O(h^2) \pr \label{how.1}
\end{eqnarray}
%%%%%%%%%%%%%%%%%%%%%%%%%%%%%%
Changing the Greek suffixes ($\m,\n,\la,\si$) to non-negative integers, say, 
(1,2,3,4), we can get the input data of 
(termnoa, weightd2ha[\ ],d2ha[\ ][\ ][\ ]) for $R_{\la\mn\si}$ and
(termnob, weightd2hb[\ ],d2hb[\ ][\ ][\ ]) for $R^{\la\mn\si}$.

The output says
%****(how.2)%%%%%%%%%%%%%%%%%%%%
\begin{eqnarray}
2R_{\la\mn\si}\times 2R^{\la\mn\si}= -8\times\mbox{A1}+4\times\mbox{B3}
+4\times\mbox{B4}+O(h^2)\com \label{how.2}
\end{eqnarray}
%%%%%%%%%%%%%%%%%%%%%%%%%%%%%%
where 
$A1=\pl_\si\pl_\la h_\mn\cdot\pl_\si\pl_\n h_{\m\la},
B3=\pl_\m\pl_\n h_{\la\si}\cdot\pl_\m\pl_\n h_{\ls},
B4=\pl_\m\pl_\n h_{\la\si}\cdot\pl_\la\pl_\si h_{\mn}$~(Table 1,Appendix A).

\vs 1
\begin{flushleft}
Ex.2\ $R_{\mn}R^{\mn}$
\end{flushleft}

$R_\mn$\ and $R^\mn$\ is expanded as
%****(how.3)%%%%%%%%%%%%%%%%%%%%
\begin{eqnarray}
2R_{\mn} &=& 2R^{\mn}+O'(h^2) \nn\\
         &=& \pl_\m\pl_\n h_{\al\al}-\pl_\m\pl_\al h_{\al\n}
           -\pl_\n\pl_\al h_{\al\m}+\pl^2h_\mn +O(h^2) \pr \label{how.3}
\end{eqnarray}
%%%%%%%%%%%%%%%%%%%%%%%%%%%%%%
Changing the Greek suffixes ($\m,\n,\al$) to 
(1,2,3) for 
termnoa, weightd2ha[\ ] and d2ha[\ ][\ ][\ ]\ (input data for $2R_{\mn}$) and
changing ($\m,\n,\al$) to (1,2,4) for
termnob, weightd2hb[\ ] and d2hb[\ ][\ ][\ ]\ 
(input data for $2R^{\mn}$), we obtain
the the input data.

The output says
%****(how.4)%%%%%%%%%%%%%%%%%%%%
\begin{eqnarray}
2R_{\mn}\times 2R^{\mn}= 2\times\mbox{A2}+2\times\mbox{A3}-4\times\mbox{B1}
-4\times\mbox{B2}\nn\\
+1\times\mbox{C1}+1\times\mbox{C2}+2\times\mbox{C3}+O(h^2)\com \label{how.4}
\end{eqnarray}
%%%%%%%%%%%%%%%%%%%%%%%%%%%%%%
where 
$A2
=\pl_\si\pl_\la h_{\la\m}\cdot\pl_\si\pl_\n h_{\mn},
A3
=\pl_\si\pl_\la h_{\la\m}\cdot\pl_\m\pl_\n h_{\n\si},
B1
=\pl_\n\pl_\la h_{\si\si}\cdot\pl_\la\pl_\m h_{\mn},
B2
=\pl^2 h_{\la\n}\cdot\pl_\la\pl_\m h_{\mn},
C1
=\pl_\m\pl_\n h_{\la\la}\cdot\pl_\m\pl_\n h_{\si\si},
C2
=\pl^2 h_{\mn}\cdot\pl^2 h_{\mn},
C3
=\pl_\m\pl_\n h_{\la\la}\cdot\pl^2 h_{\mn},
$~(Table 1,Appendix A).

%%%%%%%%%%%%%%%%%%%%%%%%%%%%%%%%%%%%%%%%%%%%%%%%%%%%%%%%%%%%%%%%%%%
%%%%%%%%%%%%%%%%%%%%%%%%%%   SEC 6  %%%%%%%%%%%%%%%%%%%%%%%%%%%%%%%
%%%%%%%%%%%%%%%%%%%%%%%%%%%%%%%%%%%%%%%%%%%%%%%%%%%%%%%%%%%%%%%%%%%
\section{Discussion and Conclusion}

When we do the suffix contraction in usual algebraic softwares
, such as REDUCE, we cannot help expressing the suffix permutation
symmetry by explicitly writing up all possible permuted cases. 
For example, in REDUCE, Q defined in Sec.2 is coded as
\vs {0.3}
%%%%%%%%%%%%%%%%%%%%%% REDUCE PROGRAM  %%%%%%%%%%%%%%%%%%%%%%%%%%%%%%%
\begin{tabbing}
INTEGER M,N; \q\q\q \= \q\q\q\q\q\q\q \\
REAL Q;             \>                \\
FOR ALL M,N LET\q \> DDH[M,N,M,N]=Q;   \\
FOR ALL M,N LET\q \> DDH[M,N,N,M]=Q;   
\end{tabbing}
%%%%%%%%%%%%%%%%%%%%%%%%%%%%%%%%%%%%%%%%%%%%%%%%%%%%%%%%%%%%%%%%%%%%%%
\vs {0.3}
where an array DDH[M,N,L,S] expresses $\pl_\m\pl_\n h_\ls$.
It is very hard to generalize this approach to higher-power terms or to
higher-rank tensors because it directly deals with all suffixes.
The number of programming lines for the definition of each invariant
increases rapidly. 
The present proposed approach do not treat suffixes directly, but
the topology of graphs (, in other words, connectivity of suffix lines).
The good efficiency of the algorithm based on the idea has been checked
by some concrete programs. 

In this paper we have explained the new algorithm using a simple
invariants of $(\pl\pl h)^2$. Its generalization to other types can be
done in a similar way. As the order of invariants increases, additional
indices are required to discriminate between invariants (graphs). 
In \cite{II97}, we have used 
a generalized version of this program and have obtained
the weak-field expansion of 
different products of three Riemann tensors. We have treated 
$(\pl\pl h)^3$-invariants (totally 90 terms) and the four indices 
are additionally
required: 1) Vertex-Type Order, 2) Number of $\pl\pl$-vertices 
and $h$-vertices, 3) Number of crossing, 4) Disconnectivity.
Most general SO(n) tensors (including invariants) constructed from
$\pl h,\ \pl\pl h,\ \pl\pl\pl h,\cdots$ can be treated similarly.
The point in this algorithm is to find good indices to discriminate
between invariants.

The present algorithm is a new 
efficient way to do the suffix contraction using a computer.
It can be applied not only to gravitational theories but also to
general field theories as stated in Sec.1.
One of interesting applications to  physics 
is the Weyl anomaly calculation in gravitational theories. 
It is known that $(\pl\pl h)^m$-
type tensor calculation is sufficient 
for the Weyl anomaly calculation
in 2m-dim gravitational theories.
It was confirmed that Weyl anomaly in 4 dim quantum gravity is
obtained by the present program\cite{II}.

\vspace{2 cm}
\begin{center}
{\Large\bf Acknowledgement }
\end{center}
%The author thanks Prof. J.A.M. Vermaseren (NIKHEF-H,Amsterdam)
%for kindly explaining the usage of
%his original and excellent algebraic software FORM 
%in the spring of 1992 at KEK,Japan.
The author  thanks Dr. N. Ikeda (RIMS,
, Kyoto Univ.) for discussions on Weyl anomaly problem
in gravitational theories.
He also thanks Prof. J.A.M. Vermaseren (NIKHEF-H,Amsterdam)
for reading the manuscript and for some comments.
%%%%%%%%%%%%%%%%%%%%%%%%%%%%%%%%%%%%%%%%%%%%%%%%%%%%%%%%%%%%%%%%%%%%

\newpage
%%%%%%%%%%%%%%%%%%%%%%%%%%%%%%%%%%%%%%%%%%%%%%%%%%%%%%%%%%%%%%%%%%%
%%%%%%%%%%%%%%%%%%%%%%%%%%  App. A  %%%%%%%%%%%%%%%%%%%%%%%%%%%%%%%
%%%%%%%%%%%%%%%%%%%%%%%%%%%%%%%%%%%%%%%%%%%%%%%%%%%%%%%%%%%%%%%%%%%
\begin{flushleft}
{\Large\bf Appendix A.\ Graphical Representation of 
$(\pl\pl h)^2$-Invariants}
\end{flushleft}

In the weak field expansion of  n-dim Euclidean gravity
:\ $g_\mn=\del_\mn+h_\mn,|h|\ll 1$,\ 
we must generally treat  global SO(n)-tensors 
which are composed of $h_\mn,\ \pl_\al h_\mn,$\linebreak 
$\pl_\al\pl_\be h_\mn,
\cdots$. Let us focus here on those tensors which are composed only of
$\pl_\al\pl_\be h_\mn$. We introduce its graphical representation
as shown in Fig.1 in Sec.2 of the text.
Contraction of suffixes is graphically represented by connecting the dotted
lines with the same suffix. For example see Fig.2 and 3 of Sec.2

In Sec.2 we have graphically presented $\pl h$-invariants (Fig.3). They are
P=$\pl^2 h\ $,\ Q=$\pl_\m\pl_\n h_\mn$\ .
Similarly we can list up all independent $(\pl\pl h)^2$-invariants
as follows. They are grouped by the number of the suffix-loop:\ 
loop no\ =1,\ Fig.14;\ 
loop no\ =2:\ Fig.15;\ 
loop no\ =3:\ Fig.16;\ 
loop no\ =4:\ Fig.17;.

%%%%%%%%%%%%%%%%%%%%%%%% Fig.14 %%%%%%%%%%%%%%%%%%%%%%%%%%%%%%%%%
\begin{figure}
     \centerline{
\psbox[width=95mm,height=40mm]{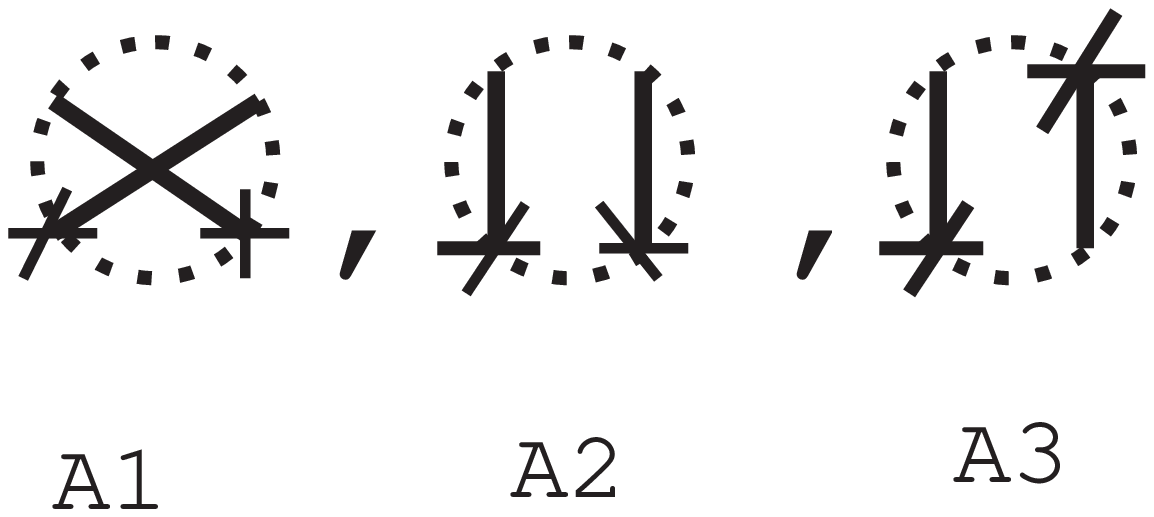}
                 }
\caption{
Graphs of $(\pl\pl h)^2$-invariants, loop no=1. 
        }
\label{fig14}
\end{figure}
%%%%%%%%%%%%%%%%%%%%%%%%%%%%%%%%%%%%%%%%%%%%%%%%%%%%%%%%%%%%%%%%%%%%%

%%%%%%%%%%%%%%%%%%%%%%%% Fig.15 %%%%%%%%%%%%%%%%%%%%%%%%%%%%%%%%%
\begin{figure}
     \centerline{
\psbox[width=160mm,height=100mm]{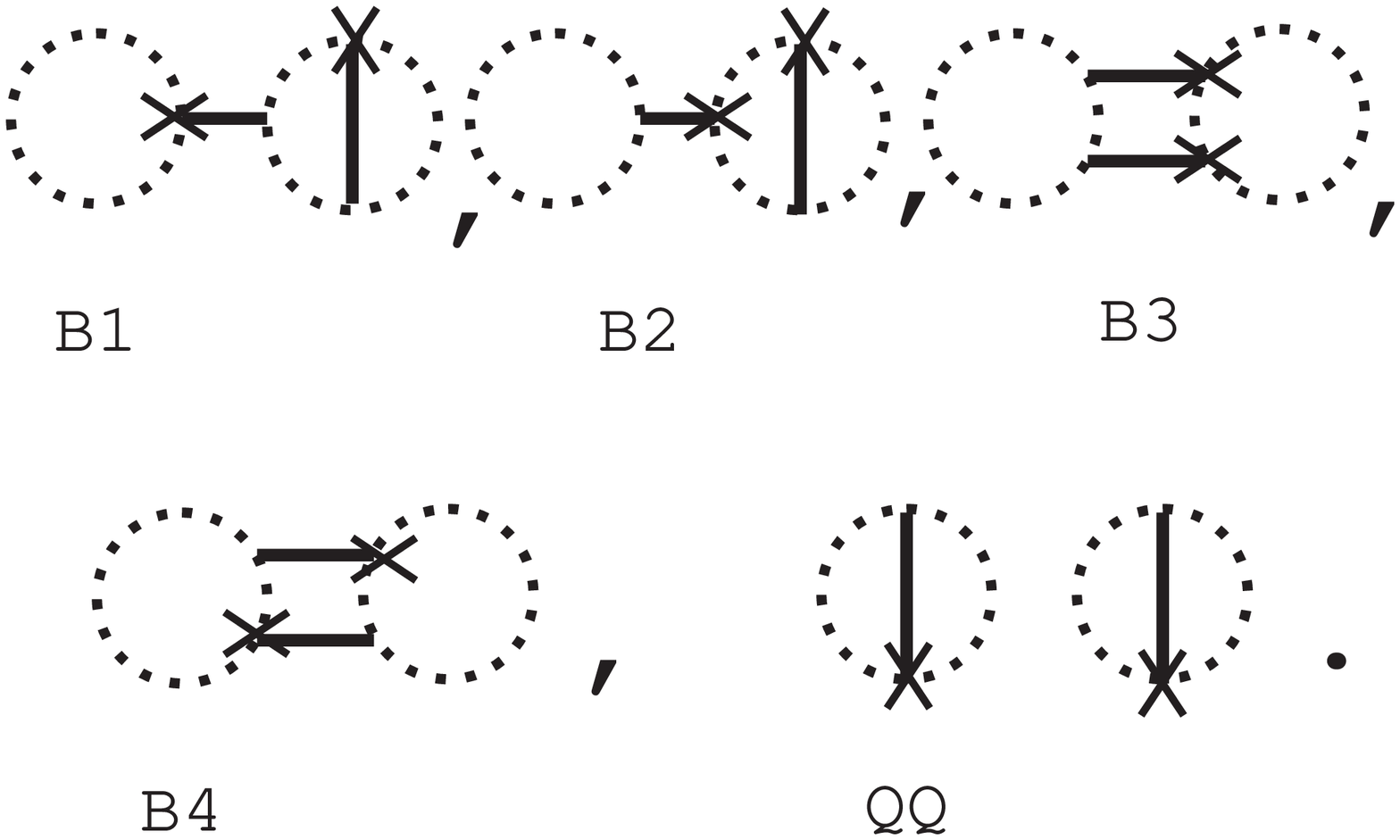}
                 }
\caption{
Graphs of $(\pl\pl h)^2$-invariants, loop no=2.
        }
\label{fig15}
\end{figure}
%%%%%%%%%%%%%%%%%%%%%%%%%%%%%%%%%%%%%%%%%%%%%%%%%%%%%%%%%%%%%%%%%%%%%

%%%%%%%%%%%%%%%%%%%%%%%% Fig.16 %%%%%%%%%%%%%%%%%%%%%%%%%%%%%%%%%
\begin{figure}
     \centerline{
\psbox[width=160mm,height=90mm]{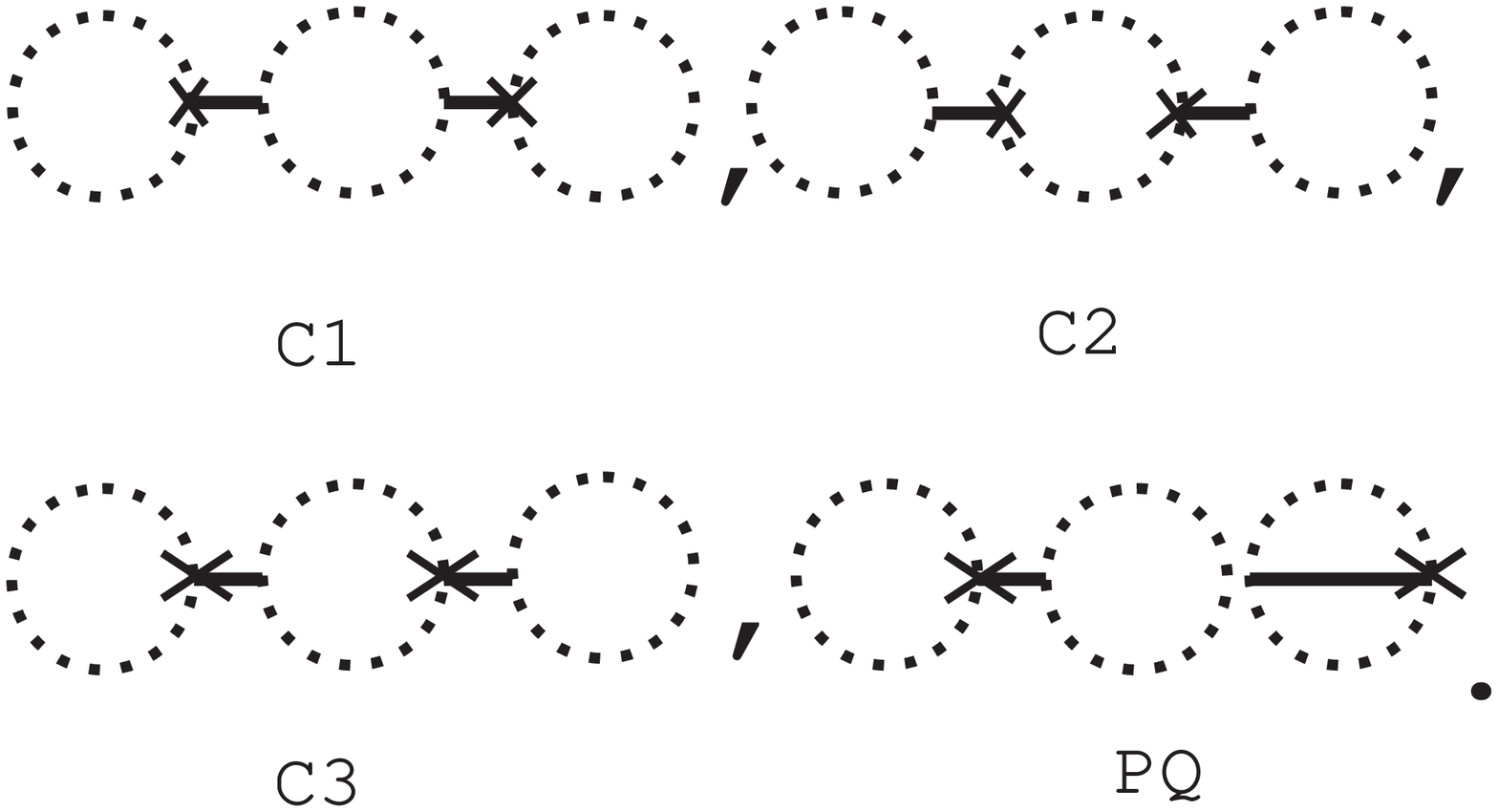}
                 }
\caption{
Graphs of $(\pl\pl h)^2$-invariants, loop no=3. 
        }
\label{fig16}
\end{figure}
%%%%%%%%%%%%%%%%%%%%%%%%%%%%%%%%%%%%%%%%%%%%%%%%%%%%%%%%%%%%%%%%%%%%%

%%%%%%%%%%%%%%%%%%%%%%%% Fig.17 %%%%%%%%%%%%%%%%%%%%%%%%%%%%%%%%%
\begin{figure}
     \centerline{
\psbox[width=105mm,height=35mm]{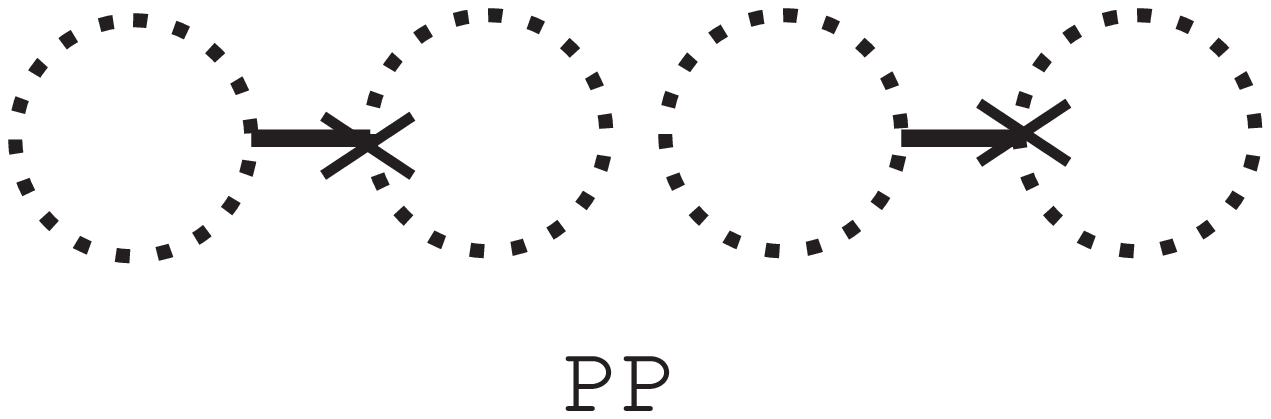}
                 }
\caption{
Graphs of $(\pl\pl h)^2$-invariants, loop no=4. 
        }
\label{fig17}
\end{figure}
%%%%%%%%%%%%%%%%%%%%%%%%%%%%%%%%%%%%%%%%%%%%%%%%%%%%%%%%%%%%%%%%%%%%%

We have totally 13 invariants (10 connected,\ 3 disconnected).
Their literal expressions are listed in Table 1 of the text(Subsec.4.4).

\newpage
%%%%%%%%%%%%%%%%%%%%%%%%%%%%%%%%%%%%%%%%%%%%%%%%%%%%%%%%%%%%%%%%%%%
%%%%%%%%%%%%%%%%%%%%%%%%%%  App. B  %%%%%%%%%%%%%%%%%%%%%%%%%%%%%%%
%%%%%%%%%%%%%%%%%%%%%%%%%%%%%%%%%%%%%%%%%%%%%%%%%%%%%%%%%%%%%%%%%%%
\begin{flushleft}
{\Large\bf Appendix B.\ Input and Output Data of Sample Calculations}
\end{flushleft}

%%%%%%%%%%%%%%%%%%%%%%%%%%% Ex.1 IN  %%%%%%%%%%%%%%%
\begin{flushleft}
Example 1,  Input Data
\end{flushleft}

\begin{flushleft}
4\\
1\\
1 2 3 4\\
-1\\
2 3 1 4\\
-1\\
1 4 3 2\\
1\\
4 3 1 2\\
4\\
1\\
1 2 3 4\\
-1\\
2 3 1 4\\
-1\\
1 4 3 2\\
1\\
4 3 1 2\\
\end{flushleft}

\vs 1
%%%%%%%%%%%%%%%%%%%%%%%%%%%%%  Ex.1 OUT   %%%%%%%%%%%%%
\begin{flushleft}
Example 1,  Output Data
\end{flushleft}

\begin{flushleft}
start of MAIN\\
$\mbox{}$\\
Input Check\\
termnoa=4\\
term0,weight=1\\
d2ha[0][0][0]=1\\
d2ha[0][0][1]=2\\
d2ha[0][1][0]=3\\
d2ha[0][1][1]=4\\
term1,weight=-1\\
d2ha[1][0][0]=2\\
d2ha[1][0][1]=3\\
d2ha[1][1][0]=1\\
d2ha[1][1][1]=4\\
term2,weight=-1\\
d2ha[2][0][0]=1\\
d2ha[2][0][1]=4\\
d2ha[2][1][0]=3\\
d2ha[2][1][1]=2\\
term3,weight=1\\
d2ha[3][0][0]=4\\
d2ha[3][0][1]=3\\
d2ha[3][1][0]=1\\
d2ha[3][1][1]=2\\
termnob=4\\
term0,weight=1\\
d2hb[0][0][0]=1\\
d2hb[0][0][1]=2\\
d2hb[0][1][0]=3\\
d2hb[0][1][1]=4\\
term1,weight=-1\\
d2hb[1][0][0]=2\\
d2hb[1][0][1]=3\\
d2hb[1][1][0]=1\\
d2hb[1][1][1]=4\\
term2,weight=-1\\
d2hb[2][0][0]=1\\
d2hb[2][0][1]=4\\
d2hb[2][1][0]=3\\
d2hb[2][1][1]=2\\
term3,weight=1\\
d2hb[3][0][0]=4\\
d2hb[3][0][1]=3\\
d2hb[3][1][0]=1\\
d2hb[3][1][1]=2\\
$\mbox{}$\\
$\mbox{}$\\
FINAL ANSWER\\
-8 A1\\
0 A2\\
0 A3\\
0 B1\\
0 B2\\
4 B3\\
4 B4\\
0 QQ\\
0 C1\\
0 C2\\
0 C3\\
0 PQ\\
0 PP\\
\end{flushleft}
\vs 1

%%%%%%%%%%%%%%%%%%%%%%%%%%% Ex.2 IN  %%%%%%%%%%%%%%%
\begin{flushleft}
Example 2,  Input Data
\end{flushleft}

\begin{flushleft}
4\\
1\\
1 2 3 3\\
-1\\
1 3 3 2\\
-1\\
2 3 3 1\\
1\\
3 3 1 2\\
4\\
1\\
1 2 4 4\\
-1\\
1 4 4 2\\
-1\\
2 4 4 1\\
1\\
4 4 1 2\\
\end{flushleft}
\vs 1

%%%%%%%%%%%%%%%%%%%%%%%%%%%%%%%  Ex.2 OUT %%%%%%%%%%%%%%
\begin{flushleft}
Example 2,  Output Data
\end{flushleft}

\begin{flushleft}
start of MAIN\\
$\mbox{}$\\
Input Check\\
termnoa=4\\
term0,weight=1\\
d2ha[0][0][0]=1\\
d2ha[0][0][1]=2\\
d2ha[0][1][0]=3\\
d2ha[0][1][1]=3\\
term1,weight=-1\\
d2ha[1][0][0]=1\\
d2ha[1][0][1]=3\\
d2ha[1][1][0]=3\\
d2ha[1][1][1]=2\\
term2,weight=-1\\
d2ha[2][0][0]=2\\
d2ha[2][0][1]=3\\
d2ha[2][1][0]=3\\
d2ha[2][1][1]=1\\
term3,weight=1\\
d2ha[3][0][0]=3\\
d2ha[3][0][1]=3\\
d2ha[3][1][0]=1\\
d2ha[3][1][1]=2\\
termnob=4\\
term0,weight=1\\
d2hb[0][0][0]=1\\
d2hb[0][0][1]=2\\
d2hb[0][1][0]=4\\
d2hb[0][1][1]=4\\
term1,weight=-1\\
d2hb[1][0][0]=1\\
d2hb[1][0][1]=4\\
d2hb[1][1][0]=4\\
d2hb[1][1][1]=2\\
term2,weight=-1\\
d2hb[2][0][0]=2\\
d2hb[2][0][1]=4\\
d2hb[2][1][0]=4\\
d2hb[2][1][1]=1\\
term3,weight=1\\
d2hb[3][0][0]=4\\
d2hb[3][0][1]=4\\
d2hb[3][1][0]=1\\
d2hb[3][1][1]=2\\
$\mbox{}$\\
$\mbox{}$\\
FINAL ANSWER\\
0 A1\\
2 A2\\
2 A3\\
-4 B1\\
-4 B2\\
0 B3\\
0 B4\\
0 QQ\\
1 C1\\
1 C2\\
2 C3\\
0 PQ\\
0 PP\\
\end{flushleft}

\vs 2
%%%%%%%%%%%%%%%%%%%%%%%%%%%%%%%%%%%%%%%%%%%%%%%%%%%%%%%%%%%%%%%%%%%%%%%%%%
%%%%%%%%%%%%%%%%%%%%%   References     %%%%%%%%%%%%%%%%%%%%%%%%%%%%%%%%%%%
%%%%%%%%%%%%%%%%%%%%%%%%%%%%%%%%%%%%%%%%%%%%%%%%%%%%%%%%%%%%%%%%%%%%%%%%%%

%%%%%%%%%%%%%%%%%%%%%%%%%%%%%
\newpage
\begin{center}
{\Large\bf Figure Captions}
\end{center}
\begin{itemize}
\item Fig.1\ 4-tensor $\pl_\m\pl_\n h_\ab$
\item Fig.2\ 2-tensors of \ (a)\ 
$\pl^2h_\ab\ ,\ (b)\ \pl_\m\pl_\n h_{\al\al}\ $ and \ (c)\ 
$\pl_\m\pl_\be h_\ab\ $
\item Fig.3\ Invariants of 
$P\equiv \pl_\m\pl_\m h_{\al\al}$\ and $Q\equiv \pl_\al\pl_\be h_{\ab}\ $.
\item Fig.4\ 4-tensor of $\pl_1\pl_2 h_{34}$.
\item Fig.5
~Invariants of $P=\pl_1\pl_1 h_{22}$~and $Q=\pl_1\pl_2 h_{12}$\ .
\item Fig.6
~8-tensor of $\pl_1\pl_2 h_{34}\cdot\pl_5\pl_6 h_{78}$.
\item Fig.7
~4-tensor of~$\pl_1\pl_2 h_{34}\cdot\pl_5\pl_6 h_{34}$.
\item Fig.8
\ Invariant of $B3\equiv\pl_1\pl_2 h_{34}\cdot\pl_1\pl_2 h_{34}$.
\item Fig.9\ \ul{loopstream}[\ ][\ ][\ ] for $B3$.
\item Fig.10\ Invariant of $B1$. \# of vertices of 0-th and 1-th loop
are 1 and 3 respectively. 0-th loop is a tadpole.
\item Fig.11\ \ul{tadpoleno} and \ul{tad}[\ ][\ ] for $C1, C2$~and $C3$.
\item Fig. 12\ Change of i (bond number) and j (vertex-type number).\nl
Arrows indicate directions of tracings.
\item Fig.13\ \ul{absdeli}[\ ] and \ul{absdelj}[\ ] for $A1, A2$~and $A3$.
\item Fig.14\ Graphs of $(\pl\pl h)^2$-invariants, loop no=1.
\item Fig.15\ Graphs of $(\pl\pl h)^2$-invariants, loop no=2.
\item Fig.16\ Graphs of $(\pl\pl h)^2$-invariants, loop no=3.
\item Fig.17\ Graphs of $(\pl\pl h)^2$-invariants, loop no=4.
\end{itemize}

\end{document}